# EPR Study of Radicals in Irradiated Ionic Liquids and Implications for the Radiation Stability of Ionic Liquid-Based Extraction Systems

Ilya A. Shkrob, *[1] Sergey D. Chemerisov,

*Chemistry Division , Argonne National Laboratory,9700 S. Cass Ave, Argonne, IL 60439*

James F. Wishart, *[2]

*Chemistry Department, Brookhaven National Laboratory, Upton, New York 11973-5000*

## Abstract

The radiation- and photo- chemistry of room temperature ionic liquids (ILs) composed of ammonium, phosphonium, pyrrolidinium, and imidazolium cations and bis(triflyl)amide, dicyanamide, and bis(oxalato)borate anions, have been studied using low-temperature Electron Paramagnetic Resonance (EPR). Several classes of radicals have been identified and related to reactions of the primary radiolytically generated electrons and holes. Large yields of terminal and penultimate *C*-centered radicals are observed in the aliphatic chains of the phosphonium, ammonium and pyrrolidinium cations, but not for imidazolium cation. This pattern can be accounted for by efficient deprotonation of a hole trapped on the cation (the radical dication) that competes with rapid charge transfer to a nearby anion. The latter leads to the formation of stable N- or O-centered radicals. The electrons either react with the protic impurity (for nonaromatic cations) yielding H atoms or the aromatic moiety (for imidazolium cations). Excitation of bis(triflyl)amide anion is shown to yield trifluoromethyl radical; the yield of this radical in radiolysis, though, is low (< 10% of the alkyl radical yield). In terms of their radiation chemistry, neat ILs appear to be intermediate between organic liquids and ionic solids. Addition of 10-40 wt% of trialkylphoshate (a common extraction agent for nuclear cycle separations) has relatively little effect on the fragmentation of the ILs. Radiation induced dealkylation of the phosphate is prominent, but the yield of the alkyl radical fragments derived from the phosphates is < 4% of the yield of the radical fragments drived from the solvent. We discuss the implication of these results for the radiation stability of nuclear cycle extraction systems based upon the IL diluents.

Authors to whom correspondence should be addressed; electronic mail: (1) shkrob@anl.gov, (2) wishart@bnl.gov

1. **INTRODUCTION**

In this study, we address the fundamental aspects of radiation chemistry and mechanisms for radiation induced degradation of room temperature ionic liquids (ILs). The latter are salts with unusually low melting points due to poor packing of bulky, oddly shaped ions. [1]

Currently, these unusual liquids are being assessed as next generation diluents for extractions in nuclear cycle separations. [2-7] The practical advantages of the ILs are due to their low volatility, high ionicity, versatile physical properties, and the possibility of integrating functional groups into the solvent. However, to be useful for the separations, these new IL and extraction systems must be chemically stable in high radiation fields of the decaying radionuclides, at least in comparison to presently-used separations systems. Not only should the solvent be resistant to highly oxidizing conditions (as the aqueous phase is 1-2 mol/dm$^3$ nitric acid), it should withstand a radiation dose that is equivalent to 1-40 Mrad (1 rad = $10^{-2}$ Gy) per cycle. [8,9] The current UREX and UREX+ extraction processes are based on normal alkanes (dodecane) as a diluent and alkylphosphates (such as tributylphosphate) as uranyl extractant. These systems exhibit relatively little radiolytic degradation of the solvent, but substantial damage to the extractant. [4-8] Dissociative electron attachment (DEA) [10,11] or energy transfer [12] to the alkyl phosphate lead to dealkylation of the phosphate esters and the formation of dialkylphosporic acids that have high affinities for radioactive lanthanides, such as Nd$^{3+}$; [9] the extraction of these cations causes radioactive contamination of the organic phase and requires costly recycling of the extractant. [8,9] From the standpoint of radiation stability, a prospective extraction system should fulfill several criteria. Not only should the absolute amount of radiation damage to the solvent (via polymerization, volatilization, and loss of fragments to the aqueous phase) be low; an equally important requirement is that the yield of a few radiolytic products capable of binding the metal ions should be negligible. Ideally, the solvent should take the brunt of radiation damage without compromising its performance as diluent while fully protecting the extractant. Alkane-based extraction systems are not optimum in this sense, as radiation generates species (electrons, holes, and excited states) that are very reactive towards the phosphate extractants, and this results in rapid deterioration of the extractant. [8-12] By contrast, halocarbon solvents provide excellent protection of the extractant, serving as scavengers of the radiolytically generated electrons and efficient quenchers of the excited states, [13] but are unacceptable for environmental (toxicity) and technological (corrosion from Cl$^-$ released into the raffinate) reasons.

A switch to ILs could thus provide an opportunity both for improving the separations efficiency and for readdressing the issue of radiation stability by finding a solvent system that *actively* protects the extractant. This is possible because ion constituents of ILs are better electron and hole traps and quenchers of the excited states than the alkylphosphate extractant. However, a greater degree of protection of the extractant means greater damage to the solvent. The latter is perhaps unavoidable for complex organic solvents. Optimization of the extraction systems involves balancing the costs of the diluent and the extractant, the separations efficiency and the radiation resistance. Rational approach to this optimizations requires detailed understanding of the mechanisms for radiation damage in several classes of ILs. As immiscibility of the ILs with acidic solutions and the suppression of cation exchange [14] are important for the extraction, the ILs of interest typically consist of large, highly hydrophobic cations (such as quaternized ammonium and phosphonium cations with aliphatic chains with 4-14 carbons) and anions that are not hydrolysable and do not chelate metal cations (such as $NTf_2^-$, $CF_3SO_3^-$, $DCA^-$ and, to a lesser degree, $PF_6^-$ and $BF_4^-$). The chemical structures for some of these ions are shown in Figure 1.

The radiation stability of several classes of ILs has already been addressed using pulse radiolysis [15-27] and product analysis. [28,29] The bulk of these studies has been on ILs composed of 1-methyl-3-alkyl-imidazolium cations ($C_n mim^+$, where *n* is the carbon number of the aliphatic chain) and bis(triflyl)imide anions ($NTf_2^-$), although salts based on Cl$^-$, $NO_3^-$, [28] and $PF_6^-$ [15,16,29] anions and quaternary ammonium ($R_4N^+$), pyrrolidinium, [25-27] and pyridinum [20] cations were also studied. Product analyses [28] suggest that $PF_6^-$ anion undergoes DEA yielding •$PF_5^-$ radical and F$^-$ (as known from EPR studies) [30] whereas $NTf_2^-$ might undergo scission of the C-S and the S-N bonds. Products of cross recombination of these radicals and C-centered radicals derived from $C_n mim^+$, by loss of hydrogen in 2-position in the ring have been observed. Other products were speculated to be formed by recombination of this –H radical with an alkyl radical in the side chain or N-centered radicals generated via the loss of the aliphatic chain. The formation of free •$CF_3$ radicals has also been proposed by Neta *et al.* [18,20] who observed radical oxidation reactions consistent with the presence of $CF_3O_2$• radicals in $O_2$-saturated IL solutions; [20] these authors also observed a 405 nm band in pyrene solutions that was attributed to the adduct of •$CF_3$ to pyrene. [18]

Pulse radiolysis studies indicate that in ILs composed of aromatic cations (such as imidazolium and pyridinium) the electrons rapidly attach to the ring to form the corresponding neutral radicals [15,16,20] that strongly absorb at 300-400 nm. By contrast, in tetraalkylammonium salts, the excess electron is trapped as a metastable F-center (the "solvated electron") exhibiting a broad absorption band in the near infrared, [23-27,31,32] suggesting that the electron is solvated by the alkyl chains of the cations. By analogy to the F-centers in ionic crystals, one can visualize these solvated electrons as occupying *s*-orbitals in anion vacancies, although functional groups present on the ILs may result in specific solvation interactions. [24] In a few hundred nanoseconds, these electrons react with adventitious protic/acidic impurities typically present in the IL and yield mobile H atoms, which can be observed indirectly using transient absorption spectroscopy via reactions with aromatic solutes to form H-adducts that have characteristic visible absorption spectra. [21,23] (In extremely pure and dry aliphatic cation ionic liquids, we have observed the solvated electron to decay on a slower timescale with complex kinetics suggestive of geminate recombination, however, such conditions are not obtained within separations processing environments that are the focus of our present interest.)

Apart from these direct observations, Neta and co-workers [16-22] postulated several primary processes that they were not able to observe directly. Specifically, they suggested that both anions ($A^-$) and cations ($C^+$) might trap holes (electron deficiencies) yielding neutral radicals (A•) and radical dications ($C•^{2+}$), respectively. The former radicals (such as sulfonamidyl N-centered radical, (•$NTf_2$) and electronically excited $NTf_2^-$ anions were thought to fragment, yielding the •$CF_3$ radicals. The radical dications may deprotonate and yield C-centered (alkyl) radicals. [20] These mechanisms may, in principle, account for the fragmentation patterns deduced from the product analyses. The main problem with the latter is that it is difficult to distinguish between primary and secondary radical chemistry.

These suggested pathways were based on the conceptual picture of IL as an organic solvent. Another reference system would be an ionic crystal composed of derivatized ammonium cations and inorganic anions. In ammonium perchlorate (R=H) [33] and tetramethyl ammonium (R=Me) chloride, [34] the main product of radiolysis is •$NR_3^+$ radical cation that is generated via homolysis of N-H or N-C bonds in the excited state of the cation. In halide crystals, the hole is trapped as •$X_2^-$ species in which there is a 3-electron, two-center σσ* bond between the halide atom and the lattice anion. One can envision that this chemicolligation also occurs for the •A radicals derived from $NTf_2^-$ and $DCA^-$. Only experiment can decide what the radiation chemistry of the ILs might resemble more: that of the ionic crystals or that of organic liquids.

To elucidate the reaction pathways, we took a different approach from these previous studies. The radicals were generated in low-temperature frozen vitreous ILs using 3 MeV electrons and 248 nm photons, and the metastable radicals were observed using continuous wave X band (cw) Electron Paramagnetic Resonance (EPR). The radicals became mobile and decayed at 180-200 K. Due to high rigidity of the IL glasses, the rotations of functional groups are frequently inhibited up to this temperature, resulting in broad, anisotropic lines and relatively low spectral resolution. Despite these complications, we were able to identify several families of fragment radicals and establish the main pathways leading to these radicals. Our study supports some of the previously suggested mechanisms and provides new mechanistic insight into the radiolytic degradation of the ILs. This insight was subsequently used to assess the radiation stability of trialkylphosphates in these ILs. Although the dealkylation of the extractant was observed, presumably via DEA, the relative yield of the alkyl radicals derived from the solute was 1-4% of the yield of the radicals derived from the solvent, even in the ILs containing 40 wt% of the extractant. These observations indicate remarkable radiation stability of the alkylphosphate extractants in this new class of diluents.

## 2. EXPERIMENTAL

*Materials:* Ionic liquids/solids were composed of the following cations: *N,N,N*-trihexyl-*N*-tetradecylphosphonium ($Hx_3TdP^+$), methyl-tri-*n*-butylammonium ($MB_3N^+$), trimethylammonium ($Me_3NH^+$), triethylammonium ($Et_3NH^+$) and tetrabutylammonium ($Bu_4N^+$), *N*-butyl-*N*-methyl pyrrolidinium ($P_{14}^+$), and *N*-methyl-*N*-decylimidazolium ($C_{10}mim^+$). Some of these ions are shown in Figure 1. The sample of $MB_3N^+$ $NTf_2^-$ was kindly provided by Dr. P. Neta of NIST and prepared as described in ref. 22. The sample of $C_{10}mim^+$ $NTf_2^-$ was provided by M. Dietz and P. G. Rickert of ANL. $P_{14}^+$ $NTf_2^-$ was prepared as described in ref. 36. $Et_3NH^+$ $NTf_2^-$ was prepared by mixing aqueous solutions containing equal amounts of triethylammonium chloride and $Li^+$ $NTf_2^-$ (Aldrich) which resulted in phase separation to produce the crude IL. The IL phase was washed repeatedly with small quantities of deionized water to remove LiCl until the rinsate tested negative for chloride ion using a silver test. $P_{14}^+$ $DCA^-$ was purchased from EMD Chemicals in "high purity" grade and used as received. Bis(oxalato)borates (BOB) of $P_{14}^+$ and $Hx_3TdP^+$ were purchased from EMD Chemicals in "for synthesis" grade and dried for 2 hours in vacuum at 50 °C before use. $Hx_3TdP^+$ $NTf_2^-$ was prepared from the aqueous metathesis of $Hx_3TdP^+$ $Cl^-$ (Cytec) and $Li^+$ $NTf_2^-$. Briefly, 5.29 g (10.2 mmol) $Hx_3TdP^+$ $Cl^-$ was suspended above 50 mL of water with stirring, while 3.02 g $Li^+$ $NTf_2^-$ (10.5 mmol) dissolved in 1.1 mL water was added. Cloudy white beads of $Hx_3TdP^+$ $NTf_2^-$ (density = 1.06 g/cm$^3$, yield ~80%) began to condense and drop to the bottom of the flask. The mixture was stirred overnight, the aqueous phase was decanted, and the ionic liquid was washed four times with 10 mL water to remove residual LiCl until the rinsate tested negative for chloride ion using a silver test, after which it was dried on a rotary evaporator, followed by drying in a vacuum oven. Tetrabutylammonium $NTf_2^-$ was prepared in a similar fashion from aqueous solutions of $Bu_4N^+$ $Cl^-$ and $Li^+$ $NTf_2^-$, except that the product was a solid that was recovered by filtration, washed with water and dried in the vacuum oven.

*Irradiation and EPR spectroscopy:* Liquid samples were placed in 4 mm diameter Suprasil tubes, degassed in several freeze-thaw cycles, sealed in vacuum, and frozen by rapid immersion into liquid nitrogen. With the exception of triethylammonium salt, room temperature ILs formed high-quality, transparent glasses; the $Et_3NH^+$ $NTf_2^-$ (which is liquid at room temperature) crystallized on cooling. In some experiments, the ILs were doped by $d_{10}$-anthracene that served both as a charge scavenger and a photosensitizer (the peak of the absorption of the anthracene is at 350 nm, whereas the ILs are transparent to the excitation light). [35]

The samples were irradiated at 77 K using either (i) 55 ns fwhm pulses of 3 MeV electrons from a Van de Graaff accelerator at Argonne (the typical dose was about 0.15 Mrad; the repetition rate was 20-30 Hz, the average current was ca. 0.6 µA), (ii) 248 nm light from a Lambda Physik LPX 120i KrF excimer laser (15 ns fwhm, 20 mJ per pulse, 4 Hz; 5-30 min), and (iii) the cw light from a 60 W Xe arc lamp that was filtered through a water and a 300-420 nm glass bandpass filter. Following the irradiation, the samples were transferred into the resonator of a 9.44 GHz Bruker ESP300E spectrometer equipped with an Oxford Instruments CF935 cryostat. The modulation frequency was 100 KHz. First-derivative EPR spectra were taken at several temperatures, microwave power levels (typically, 0.2 and 20 mW), and modulation fields (typically 1 or 10 G). The spectra given below were normalized in order to facilitate the comparison between the samples. Electron beam irradiation of suprasil yields H atoms that are trapped in the $SiO_2$:OH glass; these H atoms reveal themselves as a 500 G doublet of narrow lines superimposed on the EPR spectra from the sample. UV laser irradiation of the sample tubes under the conditions of our experiment did not result in such H atom signals. Another EPR signal that originated from the sample tube was $E_\gamma'$ defect (a Si dangling bond) whose narrow line is seen at the center of the EPR spectra.

Calculations of the geometry and magnetic parameters of the radicals were carried out using density functional theory (DFT) with B3LYP functional [37] and 6-31+G** basis set from Gaussian '98; [38] this method is most commonly used for such calculations. [39,40] In the following *a* denotes the isotropic hyperfine coupling constant (hfcc) and *b* denotes the largest (absolute) principal value of ***B***, which is the anisotropic part of the hfc tensor (these hfcc tensors are mostly axial and we chose the axis z to be associated with the long axis). These hyperfine constants and magnetic fields are given in units of Gauss (1 G = $10^{-4}$ T). The simulations of EPR spectra were carried out using the calculated hfc tensors assuming isotropic *g*-tensor (which is a gross simplification for some of the species involved, such as trifluoromethyl) using the standard first-order perturbation theory; the orientation averaging was carried out using the Monte Carlo code. The spectra were then convoluted with a Gaussian (of width $\Delta B$ to simulate the effects of inhomogeneous broadening). To save space, some of the experimental and simulated spectra are placed in the Supplement (these figures have the designator "S" (e.g., Figure 1S). The optimized geometries of the radicals and calculated hfcc tensors are also placed there.

## 3. RESULTS

### 3.1. Bis(triflyl)amide salts.

The EPR spectra of radiolyzed and photolyzed frozen ILs are rather complex, suggesting multiplicity of reaction paths leading to fragmentation reactions. The types of radicals generated are the same for photolysis and radiolysis, but the yields are different. This allows one to use the combination of the two excitation methods in order to recognize the groups of EPR signals from these different radicals.

Irradiation of the samples with 248 nm light (where the cations have strong absorption bands) results in efficient photoionization. During the low-temperature photolysis, vitreous samples yield bright luminescence in the blue, and green phosphorescence with a lifetime of a few seconds is also seen. These samples also exhibit very bright thermoluminescence and radiothermoluminescence, suggesting the formation of $S_1$ and $T_1$ states of $C^+$ by recombination of trapped charges. The likely mechanism for the photoionization is biphotonic excitation of $C^+$ (by our relatively long laser pulses) that is mediated by their $S_1$ and $T_1$ states and also by secondary photolysis of their long-lived $T_1$ states. The efficiency of photoionization by cw light from a Xe arc lamp was quite low, suggesting that the former mechanism prevails. Thus, 248 nm photolysis provides a convenient way of generating radical products via the low-energy ionization channel. By contrast, electron beam radiolysis yields the products that are generated by direct excitation of the ion constituents to highly energetic, autoionizing excited states from which fragmentation is more facile.

The strongest EPR signals in both photo- and radiolyzed samples are from C-centered alkyl radicals (Figure 2a). The spectra of these radicals are poorly resolved due to hindered chain rotations and these signals are superimposed on a narrow singlet with peak-to-peak ($\Delta B_{pp}$) separation of ca. 15 G. The latter feature is observed in all IL samples considered in this section and it can be seen unobstructed by warming the samples above 180 K, when the alkyl radicals disappear from the EPR spectra (Figure 2b). This radical must be derived from $NTf_2^-$ anion, as it is present in the ILs consisting of this anion, but it is not observed in the ILs consisting of $DCA^-$ and $BOB^-$ anions (section 3.2). This radical cannot be a C-centered radical or an F-center, and the yield of this radical does not correlate with the yield of $•CF_3$ radical (which excludes the $•SO_2NTf^-$ or $•NTf^-$ radicals).

In the EPR spectra of photolyzed $C_{10}mim^+$ and $Et_3NH^+$ salts one can also discern a weaker 67 G doublet superimposed onto this singlet (Figure 2a), which is the same for both of these salts (despite the marked difference in the cation structure). These lines are indicated by open circles in the upper two traces in Figure 2a. For the other two salts in Figure 2, this doublet cannot be seen as it is superimposed on a much stronger signal from the alkyl radicals (see below). Very similar EPR spectra were observed in electron-irradiated polycrystalline $Li^+ NTf_2^-$ (shown in Figures 1S and 2S) and $Me_3NH^+ NTf_2^-$ (shown in Figure 3S). Figure 3 shows the comparison between these two spectra observed at 50 K. For $Li^+ NTf_2^-$, the central line of this radical is better resolved at 130-165 K (Figure 1S), showing the sidebands (indicated by dashed arrows in Figure 3) with the separation of 12 G. This spectrum is superimposed on a weaker and broader signal (a doublet of septets) from an unidentified radical with hf couplings of ca. 17 G in the six fluorine atoms (possibly, from a radical pair involving lithium-7). For $Me_3NH^+ NTf_2^-$ (Figure 3S), the three-line spectrum is superimposed on a weaker pattern from $•CF_3$ radical (see below). The invariance of this spectrum with the cation structure and the change observed by the replacement of the anion indicates that the radical is derived from the $NTf_2^-$ anion.

The combination of a singlet (with poorly resolved sidebands) and a doublet (indicated by solid arrows in Figure 3) would be expected for the $•NTf_2$ radical shown at the top of Figure 3: the doublet corresponds to the extreme transitions for magnetic field aligned with the N 2p orbital of this sulfonamidyl radical (Figure 1S). The central singlet is relatively narrow as the coupling to $^{19}F$ (spin-1/2 nucleus) in the two trifluoromethyl groups is weak. Our DFT calculations suggest that for the six fluorides the $|a(^{19}F)|<1.4$ G; the estimates for $^{14}N$ (spin-1) nucleus are $a=+12$ G and $b=+24$ G, which results in the lines corresponding to the extreme transition that are separated by ca. 62 G (vs. experimental 67 G in the ILs and 82 G in Li$^+$ NTf$_2^-$). The simulated spectrum is shown in Figure 2S; the exact profile of the sidebands around the central resonance depends on the averaging of hfcc tensors for $^{19}$F in the trifluoromethyl groups . The typical estimates of isotropic hfcc for sulfonamidyl (-N•-SO$_2$-) radicals are 13-13.4 G (see ref. 41, v. II/17c, section 5.7); the anisotropic hfcc's has not been reported) for the closest analogs of the •NTf$_2$ radical, •N(SO$_2$F)$_2$ [42] and •N(SO$_3$)$_2^{2-}$, [43] $a(^{14}$N$)$=8.3 G and 13.3 G, respectively (and, for the latter radical, $b(^{14}$N$)$=24.6 G). We thus identify these features with the •NTf$_2$ radical hypothesized by Skrzypzak and Neta. [18] At higher temperature, the narrowing of the central line and slow rotation of the trifluoromethyl groups (resulting in the broadening of the outer lines) conspire to produce the singlet shown in Figure 2b (the 200 K trace therein and trace (ii) in Figure 1S). Below 100 K, the rotation is completely arrested and weak extreme transitions are seen, but the lines are inhomogeneously broadened (Figure 2a and trace (iii) in Figure 1S). Figure 3S(a) illustrates this matrix effect on the EPR spectrum of the •NTf$_2$ radical, by comparing EPR spectra obtained in (i) radiolyzed polycrystalline sample and (ii) photolyzed quenched melt of Me$_3$NH$^+$ NTf$_2^-$. While the former sample exhibits the EPR spectrum (also shown in Figure 3) in which the 67 G doublet is well resolved, in the latter sample this doublet is poorly resolved (just like the doublets shown in Figure 2a).

The observation of the sulfonamidyl radical in photoionization of C$^+$ suggests that the species is a hole trapped by A$^-$. Our results suggest that this N-centered radical is stable towards fragmentation of the C-S bond; we observed large yield of this radical even in the samples were the yield of •CF$_3$ radical was very low (see below).

Another radical that is observed in all of the ILs consisting of NTf$_2^-$, both in photolysis and radiolysis, is •CF$_3$ radical (Figure 4). The yield of this radical in radiolysis, as estimated by integration of the EPR spectra is only 5-10% of the yield of the alkyl radicals (the lowest trace in Figure 5 and simulations shown in Figure 4S convey the relative magnitude of these two groups of resonance signals). The yield of •CF$_3$ radical in laser photolysis is 10-20 times lower than in radiolysis (vs. the yield of the alkyl radicals), which suggests that this radical originates through electronic excitation of the NTf$_2^-$ anion. Addition of 10 wt% of electron scavengers, such as CH$_3$I and EtBr, considerably decreases the yield of this •CF$_3$ radical (> 5 times). Scavenging of the electrons and quenching of the excited states of NTf$_2^-$ by these dopants preclude the fragmentation of the dissociative excited states of the anion. The EPR spectrum of •CF$_3$ radical (which is superimposed on the much stronger signals from the N- and C-centered radicals) can be seen in all EPR spectra shown in Figure 4a. The EPR pattern is rather complex due to the strong anisotropy of the *g*-tensor and the three hfc **A**-tensors for $^{19}$F nuclei in the trifluoromethyl. [42-47] Ref. 45 contains detailed theoretical analysis of such EPR spectra, including the second order effects. As our intent was to identify the radicals rather than account for the intricacies of their EPR spectra, we have not attempted such analysis.

The two extreme transitions (AA') separated by $3A_\parallel^{eff} \approx 692$ G in Figure 3b (for radiolyzed Et$_3$NH$^+$ NTf$_2^-$) correspond to the alignment of the magnetic field *B* with the axis of symmetry for this $C_{3v}$ symmetrical radical. [45] The two groups of doublets (BB' and CC') are typically observed in the EPR spectra of trifluoromethyl above 10-20 K. [45,46] These doublets can be fully resolved at 130 K, due to the better averaging of the anisotropies. As $a(^{19}F)=144$ G weakly depends on the matrix, [44-47] we obtain $b_\parallel^{eff} = b(3\cos^2\alpha - 1)/2 \approx 87$ G, where α is the angle between the F 2p orbital and the symmetry axis. For a typical angle of 18°, [44,45] $b \approx 101.5$ G (vs. 119 G for •CF$_3$ in solid Kr at 4 K). Such complicated EPR spectra from noninverting, nonrotating •CF$_3$ radicals have been observed only in cryogenic crystals, at 4-20 K. In frozen ILs such features, with very little or no dynamic averaging, can be observed at temperatures as high as 170 K. This suggests extreme rigidity of the IL glasses. The same is suggested by incomplete averaging of the anisotropies in the N- and C-centered radicals.

As stated above, though we observed strong EPR signals from H atoms in radiolyzed samples (accounting for < 0.5% of the total radical yield), these signals might originate from the sample tubes. As the weaker, saturable EPR signals from these H atoms at 77 K are also observed in laser photolysis, and such signals do not occur in ILs doped by the electron scavengers such as anthracene, nor in imidazolium salts, we believe that some of these trapped H atoms were indeed formed in the radiolyzed samples. Stabilization of the H atom in an organic glass at 77-120 K is unprecedented: typically, the H atoms promptly decay (as these light atoms can readily tunnel) by H abstraction and/or addition to double bonds. The unexpected stability (albeit at low yield) of the H atom suggests that the trapping site is formed by the (inert) $NTf_2^-$ anions, i.e., this trapping site is a cation vacancy. As the likely mechanism for the production of these trapped H atoms is a reaction of the excess electron (solvated or dry) with some residual protic centers (such as impurity $R_3NH^+$ cations [23,24] that are H-bonded to $NTf_2^-$); such trapping seems likely, as the H atom is produced next to the anion in a location where a cation existed before dissociative electron attachment. The reaction with residual $H_3O^+$ seems unlikely, as no trapped D atoms were observed in the ILs that were shaken with $D_2O$, suggesting that the $e^-$-trapping center cannot be H/D exchanged in this fashion. Note that only when the parent anion is situated next to the cation vacancy can the trapped H atom be metastable; otherwise it would rapidly react with hydrocarbon moieties. The stability of a subset of the H atoms is therefore indicative of the formation of the cation vacancies; in this respect the ILs resemble ionic solids. To stabilize the H atoms, this trapping mechanism requires rigidity of the matrix. As we have seen, this is indeed the case in the frozen IL glasses.

Apart from the H atoms, we did not observe other radicals that can be traced to the electron trapping centers in the ammonium and pyrrolidinium salts. In particular, no trapped electron species (the F-centers) [23-27,32,33] or $•CF_2SO_2NSO_2CF_3$ radicals were observed. Short-lived F-centers were observed [23,24,26] in room-temperature ILs; either these centers are short-lived even at low temperature (decaying by reaction with the protons) or their (microwave saturable) signals are masked by other radicals. The $•CF_2^-$ radical (the product of DEA involving the triflyl group of the anion) does not appear to be formed. The loss of F with the formation of $CF_3CO_2^-$ was observed in gamma radiolysis of $NH_4^+ CF_3CO_2^-$,[47] and such reactions are fairly typical for other fluorinated compounds (such as $CF_2(CONH_2)_2$, $(CF_2CO_2^-)_2$).[44] In radiolysis of crystalline $CF_3CONH_2$[46] and $NH_4^+ CF_3CO_2^-$[47] at 77 K, the yields of the $•CF_3$ and -F radicals were comparable.

The two weak lines shown by arrows in Figure 4b might be, in principle, be attributable to the extreme spin transitions arising from such $•CF_2^-$ radicals,[44] although the scaling of these signals with the features of $•CF_3$ (as a function of dose, exposure, and temperature) argues that these are associated with the latter radical (some weak additional lines may appear due to the anisotropy of the *g*-tensor). Perhaps the DEA is suppressed due to the availability of deeper electron traps in the solid.

One of the possible decay routes for the excess electron is DEA to quaternized ammonium cations with the formation of the amine and dealkylation. The dealkylation or the loss of a hydrogen atom (with the formation of stable •$NR_3^+$ radical cations) is also possible via direct excitation and fragmentation of •$NR_4^+$ cations that is known to occur in the ionic crystals. [33,34] Our results argue against the occurrence of this reaction in irradiated ILs. The •$NR_3^+$ radical cation would exhibit large $H_\beta$ splittings in their aliphatic chains (ca. 26.7 G for •$NMe_3^+$ in tetramethylamine chloride; [34] the resulting pattern from the six magnetically equivalent protons is not seen in the EPR spectra. The only system examined in which such a radical cation was generated by radiolysis was polycrystalline $Me_3NH^+$ $NTf_2^-$, where the octet of lines separated by 28.9 G was seen even at 300 K (Figure 3S(b)). As this pattern is not observed below 120 K and it appears at the same temperature at which the •$NTf_2$ radical disappears (around 150 K), it is likely that this species is produced by H atom abstraction from the parent trimethylammonium cation by the •$NTf_2$ radical rather than the homolysis of the N-H bond in the excited state of the parent cation. No such species was observed in frozen liquid $Et_3NH^+$ $NTf_2^-$ and in molten $Me_3NH^+$ $NTf_2^-$ (quenched in liquid nitrogen). Apparently, this secondary reaction occurred only in crystals.

The dealkylation via DEA involving the parent cations, $C^+$, can only result in the formation of terminal alkyl radicals, such as ethyl radical for $Et_3NH^+$. By contrast, proton transfer reactions involving the hole trapped at the cation site would produce $Et_2NH^+$•$CHCH_3$ and $Et_2NH^+CH_2CH_2$• radicals. As the absolute isotropic hfcc's for $H_\alpha$ and $H_\beta$ protons in the alkyl radicals are close, the EPR spectra from the ethyl radicals (five protons) would yield qualitatively different pattern than these radicals (four protons). This can be demonstrated by direct simulation of EPR spectra using the hfcc's reported in the literature (Figure 6S) and by generation of the ethyl radical in the matrix via photolysis of the samples containing EtBr (see the upper trace in Figure 5) The EPR spectrum observed from irradiated $Et_3NH^+$ $NTf_2^-$ crystals at 77 K looks rather complex due to poor rotational averaging, but warming the sample to 180 K removes most of the extra features by permitting the rotational dynamics (Figure 5), and the EPR spectrum of the alkyl radical can be clearly seen. This spectrum is undoubtedly from a four-proton alkyl radical rather than the ethyl, so the dealkylation is excluded by our EPR results. This reaction is probably too slow to compete with other channels of electron decay, such as the reaction with protic impurity. Our data, however, do not exclude the possibility of generation of H atoms via DEA to $Et_3NH^+$.

While the DEA to tetraalkylammonium does not seem to occur, the formation of a neutral radical by electron attachment to imidazolium is fully expected to occur, as these radicals have been observed in pulse radiolysis of imidazolium salts, identified by their 320 nm absorption band. [15,16] In the $C_{10}$mim• radical (Figures 7S(a) and 8S(a)) the spin density is divided between N(1) and N(3) nitrogens and C(2) carbon in the imidazolium ring. Our DFT calculation indicates that the proton coupling for the C(2)-H should be ca 17.4 G, and proton couplings for protons in $\beta$ position to N 2p orbitals should also be considerable (6.7 G), whereas $^{14}$N couplings are relatively small (about 3.5 G). The EPR spectrum, given large broadening, would appear as a singlet with a peak-to peak splitting of ca. 44 G (Figure 8S, trace (i) in panel (b)). The EPR spectra observed in radiolyzed $C_{10}mim^+ NTf_2^-$ do exhibit an unresolved, broad singlet (Figure 6a) that can be interpreted this way (with $\Delta B_{pp}$ of ca. 40 G). Lack of spectral resolution and the superimposed signals from other radicals (including •$CF_3$) prevented us from positive identification of this radical. We may, however, conclude that electron trapping by imidazolium cations is not inconsistent with our magnetic resonance results.

We turn now to alkyl radicals that are the prevalent feature in the EPR spectra of radiolyzed (and even to a greater degree) photolyzed ILs. As stated above, for the triethylammonium salt, the alkyl radical cannot be the ethyl radical. Previous EPR studies of $\gamma$–irradiated crystalline alkylamine chlorides, [48] indicated that only terminal $Et_2NH^+CH_2CH_2$• radicals (with $a(H_\alpha)$=28 G and $a(H_\beta)$=18-20 G) were formed in $Et_3NH^+Cl^-$. This result was in line with the fragmentation patterns for other mono- and diethylammonium chlorides [48] indicating that that only terminal radicals were formed in $\gamma$–radiolysis. Our DFT calculations suggest that isotropic hf constant on $^{14}$N for the interior radical ($Et_2NH^+CH•CH_3$) is ca. –2.6 G and the splitting on $H_\beta$(NH) is ca. 0.8 G, so in both of the alkyl radicals, the terminal and the interior, the EPR spectrum is due to hfcc's to $\alpha$ and $\beta$ protons in the aliphatic chain. Given that the isotropic hfcc's are very close, it is impossible to distinguish the two types of radicals in this case. However, for $MB_3N^+$, such a distinction can be made because $-CH_2CH_2CH_2\bullet$ (radical (i) in Figure 7) and $-CH_2CH\bullet CH_3$ radicals (radical (ii) in Figure 7) (that have comparable isotropic hfcc's for $\alpha$ and $\beta$ protons) have different numbers of strongly coupled protons (four for the terminal, six for the penultimate, and five for the interior). Comparison with the simulated EPR spectra and the spectrum observed for *n*-propyl radical in solid butyronitrile (this radical may be generated via DEA to the nitrile) [49] in Figure 2a conclusively indicate that the alkyl radicals derived from $MB_3N^+$ are terminal. Exactly the same radicals are observed in radiolysis of $MB_3N^+$ $NTf_2^-$ glass and polycrystalline $BuN^+$ $NTf_2^-$ (Figure 9S) suggesting that the methyl group does not fragment. For photolyzed $P_{14}^+$ $NTf_2^-$, both types of radicals (the penultimate and the terminal) are present (as indicated by the outer pair of lines indicated by filled circles in Figure 2a), suggesting less selective fragmentation. In radiolyzed $P_{14}^+$ $NTf_2^-$ glass, the lines from the interior/penultimate radicals are more easily observed (Figure 6b). Radiolysis (as opposed to photolysis) of $MB_3N^+$ $NTf_2^-$ also yields these outer lines, though the relative amplitude of these signals is 2-3 times lower than in the $P_{14}^+$ (Figure 10S). Apparently, there is less bias towards the generation of terminal alkyl radicals in radiolysis as compared to photolysis.

### 3.2. *Other classes of ILs.*

In this section, we briefly examine EPR spectra obtained for other classes of ILs composed of hydrophobic ions.

Figure 11S(a) shows the comparison of EPR spectra obtained for radiolyzed (powder) $Na^+$ $DCA^-$ and photolyzed $P_{14}^+$ $DCA^-$ glass (Figure 1). The EPR spectrum observed in sodium dicyanamide is consistent with that expected for an inhomogeneously broadened lines of the $\bullet N(CN)_2$ radical simulated in Figure 11S(b) using hfcc's estimated from our DFT calculations. In this $C_{2v}$ symmetrical radical, the unpaired electron is shared by three $\pi$-orbitals in the nitrogen atoms residing in a $b_1$ SOMO, so that the atomic spin density is split 4:5:4 between the three nitrogen atoms. The B3LYP/6-31+G** calculation gave the estimates of $a=4.7$ G ($b=9.2$G) for the central $^{14}$N nucleus and $a=11.4$ G ($b=15$ G) for the two cyanide nitrogens. In photolyzed $P_{14}^+$ $DCA^-$ glass (Figure 10S(a)), this spectrum is superimposed onto the spectrum of the alkyl radicals derived from the $P_{14}^+$ cation as it is observed in photolyzed $P_{14}^+$ $NTf_2^-$. It is seen that the type of the alkyl radicals is the same in both of the ILs. Thus, the behavior of the dicyanamide and bis(triflyl)amide anions appears to be similar: both of these anions serve as efficient hole traps.

In Figure 12S, EPR spectra of photolyzed and radiolyzed $P_{14}^+$ $BOB^-$ glass are shown. These spectra are dominated by a narrow, asymmetrical singlet from an oxygen-centered radical derived from the $BOB^-$ anion. This resonance line might be composite, as the photolysis produces a narrower and more symmetrical line than the radiolysis. In the wings of this line, the spectral lines of the alkyl radicals derived from the cation are observed; the relative yields of these alkyl and O-centered radicals is 1:2.3.

In crystalline potassium and ammonium oxalates, [50] the oxalate anion traps the holes, forming $•O_2C\text{-}CO_2^-$ radical. The hole scavenging by $BOB^-$ anion, in analogy to this hole trapping reaction and the formation of boron oxygen hole centers involving tetrahedrally coordinated $BO_4^-$ centers in borate [51] and borosilicate [52] glasses would be via the dissociation of the B-O bond with the formation of the trigonal $BO_3$ unit in the $(OCO)_2BOCO_2•$ radical. The electron attachment would be dissociative, resulting in the partial or complete elimination of the oxalate anion and the formation of a boron dangling bond (B E') center (also known as boron electron center, BEC, for trigonal coordination), $(OCO)_2B•$ or $(OCO)_2B^-•O(CO)CO_2^-$, – in analogy to radiolysis of vitreous boron trioxide and alkali borate glasses. [51] The EPR spectra of radiolyzed samples of $BOB^-$ glasses do reveal a weak doublet of broad resonance lines separated by 238 G (not shown), that would be consistent with the inner $M(^{11}B)=\pm 1/2$ lines of this $^{11}B$ E' or BEC center. [51] The DFT calculation for the tentative $(OCO)_2B•$ radical gives an estimate of $a(^{11}B) =290$ G and $b(^{11}B)=21$ G.

The scavenging reaction of the solvated electron by $BOB^-$ was observed in room temperature $P_{14}^+$ $NTf_2^-$ containing $P_{14}^+$ $BOB^-$ by pulse radiolysis transient absorption spectroscopy at the BNL LEAF Facility. A rate constant of 3.1 x $10^8$ $M^{-1}$ $s^{-1}$ was obtained, close to the diffusion limit in $P_{14}^+$ $NTf_2^-$. [53]

The EPR spectra observed for irradiated $Hx_3TdP^+$ $NTf_2^-$ and $Hx_3TdP^+$ $BOB^-$ glasses are similar to those observed for ILs composed of the ammonium cations with the same anions (Figure 13S), except that in the bis(triflyl)amide salt, almost no $•CF_3$ radicals are formed. In the bis(oxalate)borate salt, the ratio between the yield of the alkyl radicals derived from the cation and the O-centered radicals derived from the anion is 1:1.6 (in 3 MeV electron radiolysis) and 1:1.1 (in 248 nm laser photolysis), respectively. It appears that the ILs consisting of large, hydrophobic phosphonium and ammonium cations have rather similar radiation chemistry.

### *3.3.  Radiation stability of trialkylphosphates in the ILs.*

As observed in the Introduction, radiation stability of the alkylphosphate extractant (of which 30 wt% is typically added to the solvent during the actinide separations) matters for nuclear cycle separations as much (if not more) than the radiation stability of the diluent. The specific concern is the dealkylation of the phosphate esters $((R'O)_3PO)$ that results in the formation of dialkyl phosphoric acid $((R'O)_2P(O)OH)$ capable of nonspecific interactions with methal ions. [8-12] While the acid residue $((R'O)_2P(O)O^-)$ generated in the course of DEA cannot be observed using EPR, the complementary alkyl radical (R'•), can be observed directly, provided that the resonance lines of this radical can be distinguished against the strong background of the radicals derived from the solvent (section 3.1). For these reasons, the use tributyl phosphate (that is commonly used as the extraction agent) is inconvenient, as the resulting butyl radicals cannot be distinguished from the radicals generated by deprotonation of (aliphatic) cations. Instead, we studied the dealkylation of trimethyl- and triethyl-phosphates, as the corresponding methyl and ethyl radicals (Figure 5S) have characteristic spectral lines that can be observed even in the presence of other alkyl radicals and these radicals can be readily generated in the IL glasses via DEA to alkyl halides, so their EPR spectra in rigid IL glasses can be obtained and used for calibration of the relative concentrations. These methyl and ethyl radicals were stable at 77 K, so their yield provided a good quantitative measure of the radiation stability of the solute with respect to the dealkylation.

Figure 8 exhibits EPR spectrum from radiolyzed $C_{10}mim^+$ $NTf_2^-$ glass containing 35 wt% trimethylphosphate (TMP). Since there is almost no alkyl radicals generated in the neat imidazolium salts (section 3.1), the four resonance lines of the methyl radical were readily observed against the broad, featureless spectrum of $C_{10}mim•$ radical. Comparison of these EPR spectra with the known EPR spectrum of methyl radical in this IL (shown in the same plot, lower trace) indicates that the yield of •CH$_3$ was ca. 2% of the C$_{10}$mim• radical. Addition of TMP (similar to other electron scavengers) also decreased the yield of •CF$_3$ radicals by a factor of 3.5. The yield of the methyl radicals in laser photolysis is considerably greater (the lower trace in Figure 8), due to the direct photoexcitation and photodissociation of the TMP by 248 nm light. Similar results were obtained for C$_{10}$mim$^+$ NTf$_2^-$ and P$_{14}^+$ BOB$^-$ glasses containing 10 wt% and 35 wt% and 20 wt% of triethylphosphate (TEP), respectively. The distinctive resonance lines from the ethyl radical are clearly seen in the photolyzed samples (Figure 14S), but barely seen in the radiolyzed ones; the yield of the ethyl radical is less than 2-3% of the total yield of the matrix radicals. For P$_{14}^+$ NTf$_2^-$ containing 14 wt% or 40 wt% TMP, the lines from the methyl radical stand out against the broader resonance lines of the alkyl radicals derived from the P$_{14}^+$ cation. The relative yield of these radicals is ca. 1% for 14 wt% TMP and 4% for 40 wt% TMP (in the latter solution, the yield of •CF$_3$ radical was 2 times lower than in neat IL). Importantly, the relative yield of the methyl radicals *decreases* with the increasing dose (the yield of dealkylation was observed to decrease with the dose in the alkane solvents, too). [8] This decrease can be used to contrast the EPR lines from the methyl radical against the underlying lines from the IL matrix, as shown in Figure 15S(b). The largest relative yield of the methyl radicals (with regards to the alkyl radical derived from the IL solvent) of ca. 8% was observed in radiolysis of 20 wt% TMP in P$_{14}^+$ BOB$^-$; however, even in this sample the methyl radical accounted for only 3% of the *total* radical yield (Figure 16S).

We conclude that although radiation-induced dealkylation of the phosphate esters does occur in these IL glasses, the relative yield of this fragmentation does not exceed 1-4% of that of the solvent, even when the concentration of the extractant is 20-40 wt%. Whereas the radiation stability of neat ILs is inferior to simple organic diluents, such as paraffins, these IL solvents can actively protect the extractant against the radiation damage. The DEA to the extractant and its subsequent dealkylation do occur, but it is a relatively minor reaction, even at the higher end of the concentration range.

4.     DISCUSSION

Recapitulating the previous section, three classes of radicals were observed in radiolyzed ILs consisting of NTf$_2^-$ anions. The radicals other than trifluoromethyl can be accounted for by assuming the following reaction scheme:

$$h^+ \bullet + C^+ \to C \bullet^{2+} \tag{1}$$

$$h^+ \bullet + A^- \to A \bullet \tag{2}$$

$$C \bullet^{2+} + A^- \to C(-H) \bullet + H^{\delta+} A^{\delta-} \quad \text{(for aliphatic)} \tag{3a}$$

$$C \bullet^{2+} + A^- \to C^+ + A \bullet \quad \text{(for aromatic)} \tag{3b}$$

$$e^-_{(t)} \bullet + C^+ \to C \bullet \quad \text{(for aromatic)} \tag{4}$$

$$e^-_{(t)} \bullet + BH^{\delta+} \cdots A^{\delta-} \to H \bullet + B + A^- \quad \text{(for aliphatic)} \tag{5}$$

$$H \bullet + V^+ \to H_t \bullet \quad \text{(for aliphatic)} \tag{6}$$

$$e^- \bullet + V^- \to e_t^- \bullet \quad \text{(for aliphatic)} \tag{7}$$

$$e^-_{(t)} \bullet + (R'O)_3 PO \to \to R' \bullet + (R'O)_2 P(O)O^- \tag{8}$$

The holes are initially trapped by either cation (reaction (1)) or anion (reaction (2)) moieties. The anion is a deeper trap for the holes and a good proton acceptor, so the radical dication ($C\bullet^{2+}$) either rapidly deprotonates (reaction (3a)) or accepts the electron from the anion (reaction (3b)). For aliphatic cations, the deprotonation route is prevalent; for aromatic cations, the charge transfer route is prevalent. The most abundant radicals in ILs that consist of aliphatic cations are alkyl radicals that are generated via the deprotonation of the radical dications. The aromatic cations also serve as electron-trapping centers; for aliphatic cations, the DEA reaction does not occur. For imidazolium cations, the alkyl radical yield is greatly reduced and the EPR spectra are dominated by the $C_{10}$mim$\bullet$ (ring) π-radical generated in reaction (4). Similar reactions have been shown to occur for other aromatic cations, such as pyridinum cations.[20] In ILs consisting of aliphatic cations, the electrons are trapped by anion vacancies (V$^-$), reaction (7). Both the free and the trapped electrons [23-27] react with hydrogen-bonded protic impurities (BH⋯A, such as R$_3$NH$^{\delta+}$⋯$^{\delta-}$O=S(O) or HOH$^{\delta+}$⋯$^{\delta-}$O=S(O) centers) or amides (reaction (5)) to generate mobile H atoms. A small fraction of these H atoms are trapped at the cation vacancy sites (V$^+$), reaction (6); the rest decay via abstraction and addition reactions involving the IL. The $NTf_2^-$ anion competes with the cations as a hole trapping site (reaction (2)), yielding relatively stable sulfonamidyl radical; analogous reactions occur for $DCA^-$ and $BOB^-$ anions. Electronic excitation of the $NTf_2^-$ anion, either direct or via charge recombination, results in its fragmentation with the formation of •$CF_3$ radicals. The yield of these radicals is relatively low as compared to the yield of radicals generated in reactions (1) to (7). The reaction of the solvated electron with protonated anion ($HNTf_2$) is not considered explicitly here because ionic liquids in contact with reprocessing systems will contain enough water to make reaction (5) dominant. In ref. 21 it was shown that the reaction of $e^-$ with $HNTf_2$ does not result in the production of H atoms – the presence of water is required. Reduction of $HNTf_2$ may result instead in DEA, and will be the subject of further studies.

The EPR spectra suggest that pulse radiolysis and photoionization of quaternized ammonium salts results in the formation of, predominantly, *terminal* and penultimate alkyl radicals derived from the longest chains. The example of triethylammonium indicates that these radicals are attached to the cation (rather than formed via dealkylation in an excited state or DEA).

These results are broadly consistent with a scenario in which such radicals are generated via reaction (3a). In crystals of alkylammonium salts, the cations typically assume $C_s$ symmetrical "letter T" configuration with fully extended, aliphatic chains pointing in the direction of the counter anions (e.g., refs. 54 and 55). When the hole is trapped by the cation, the excess positive charge localizes in such a way as to minimize Coulomb repulsion. This behavior has the earlier precedent in matrix-stabilized radical cations of *n*-alkanes, [56-60] where the maximum spin density is observed in the terminal, in-plane hydrogens. Although the singly occupying molecular σ-orbital (SOMO) spreads over the entire aliphatic chain, the hfcc's on the inner protons are negligible. Depending on the packing of the alkane molecules in a paraffin crystal, it is either these terminal or penultimate hydrogens [56-60] that are lost in proton transfer (leading to even-odd alternation in the yield of the corresponding radicals as a function of carbon number, which correlates with the alternation of crystal symmetry). [59,60] EPR studies of multiple conformers of linear and branched alkane radical cations stabilized in Freon and $SF_6$ matrices indicate that proton transfer preferentially occurs at the H site of maximum unpaired electron density.[56] For quaternized alkylammonium cations, both of these factors (molecular packing and spin localization) favor deprotonation of -CH$_2$CH$_3$ hydrogens: the spin density in the terminal and penultimate hydrogens is the largest and these sites are the closest to the proton-accepting anions.

To put this reasoning on a firmer footing, we investigated the SOMO for the radical dications using DFT methods (Figure 9 shows the isodensity maps of the SOMO for the optimized geometry, *C$_s$* symmetrical species). For Et$_3$NH•$^{2+}$ (Figure 9a) most of the atomic spin density and the excess positive charge are in the terminal methyl groups (2:1 to the interior hydrogen atoms). For methyltripropylammonium•$^{2+}$ and P$_{14}$•$^{2+}$ (Figures 9b and 9c), the maximum spin density and the excess positive charge are distributed 2:3 between the terminal and the penultimate sites. By contrast, for C$_4$mim•$^{2+}$ shown in Figure 9d (that served as a model cation for C$_{10}$mim•$^{2+}$), the two alkyl sites (that equally divide the spin density on the protons) contain only 20% of the *total* spin density, whereas 80% of the spin density resides in the π-orbitals in C(2), C(4), and C(5) atoms of the imidazolium ring. This suggests that C$_4$mim•$^{2+}$ is more homologous to aromatic radical cations than the alkane radical cations. (the pulse radiolysis study of Behar et al.[16] is in agreement with this conclusion: a species attributed to the radical dication generated by oxidation of the imidazolium cation by SO$_4$•$^-$ was observed in an aqueous solution). Thus, this radical dication should be more stable towards the proton transfer from the alkyl chains and thus decay via reaction (3b). This would account for the reduced production of alkyl radicals in the imidazolium based ILs. The charge transfer should also be facilitated by fortuitous ion packing observed in imidazolium - NTf$_2$$^-$ crystals:[61-63] the planar imidazolium ring is typically situated next to the -N-[62] or -SO2-[63] groups of the NTf$_2$$^-$ anion.

While several substitution products at the C(2) site of the imidazolium ring were observed by Berthon et al.[31] (including CF$_3$ and alkyl substitutions) and interpreted by these authors as evidence for cation fragmentation, these products can also be generated by addition of these radicals to the imidazolium ring and subsequent deprotonation. We observed no evidence for dealkylation of the imidazolium cation postulated by these workers.[29] The caveat is that according to our DFT calculations, the C-centered radicals (e.g., Figures 4S(b,c)) formed by the loss of hydrogen in the imidazolium ring would have relatively small hfcc's (see the Appendix) and exhibit EPR spectra represented by a narrow singlet with the peak-to-peak splitting of ca. 20 G (Figure 5S). This singlet might be obscured by EPR lines of other radicals, like the tentative EPR signal from the $C_{10}$mim• radical discussed in section 3.1. While such a possibility cannot be excluded, it presently appears that imidazolium cations are more stable towards fragmentation via reaction (3a) than other cations studied here.

## 5. CONCLUSION

Our EPR studies begin to provide a coherent, albeit incomplete, picture of the initial stages of radiation damage in several practically important classes of room temperature ILs. The process that yields most fragments is trapping of the hole by alkylammonium (phosphonium) cations which results in their rapid deprotonation from the terminal (and to a lesser degree) penultimate sites of their longest extended aliphatic chains. Our examination of the trapped-hole dication species qualitatively accounts for the fragmentation patterns observed. The formation of interior radicals is precluded because the spin density in the alkyl chains is maximal for the terminal and penultimate sites of the longest extended chain, from which the deprotonation occurs. That specifically these sites become predominant is the likely consequence of packing of IL glasses, as these sites are the closest to the proton-accepting anions. The H-abstraction reactions of H atoms (or other mobile radicals) with the alkyl chains would be expected to produce a pattern that reflects the relative abundance of the three proton donating sites, with the interior and penultimate radicals providing the largest contribution. The clear preference in the generation of terminal radicals is thus indicative of the deprotonation of a short-lived radical dication of the ammonium/phosphonium as the predominant mechanism for alkyl radical formation.

The deprotonation is less important for imidazolium cations (and, perhaps, other aromatic cations, such as pyridinium) as the spin density mainly resides on the aromatic ring and charge transfer to a nearby anion can take place instead. This does not mean that the imidazolium cations are protected. Quite to the contrary, the aromatic cations serve instead as electron trapping centers that generate π-radicals which initiate yet another degradation path. Hole trapping by the cations competes with the hole trapping by the anions that are deeper traps (and thus would prevail in the absence of deprotonation); the resulting species is an N-centered radical (for $NTf_2^-$ and DCA) or an O-centered radical (for $BOB^-$) that is rather stable. The electrons are trapped either by the aromatic cations or, in the ILs consisting of the aliphatic chain cations, at the adventitious protic sites (having first been localized as F-centers in anion vacancies, as indicated by pulse radiolysis studies). [16-20] The latter convert to H atoms, some of which are trapped in cation vacancies at low temperature. Recombination of direct excitation at the anion site causes fragmentation of the $NTf_2^-$ anion with the release of •$CF_3$ radicals. The $BOB^-$ anion appears to be capable of electron scavenging via DEA with the formation of a boron dangling bond radical.

In the context of the greater problem that we address (how to increase the radiation stability of extraction systems based upon an IL diluent), the sketch of the radiation chemistry drawn above is discouraging in some respects and encouraging in other respects. First, it is seen that the radiation damage to the cation is unavoidable, as the deprotonation is not precluded even by reaction (3a) that involves anions which constitute half of the liquid: if the cation does not fragment via reaction (3b), it is reduced in reaction (4); in both of these cases, reactive radicals are formed and oligomerization is possible. Second, the excitation of both ions occurs and can result in prompt fragmentation of these ions. Such reactions also cannot be precluded. Third, mobile, reactive radicals such as •$CF_3$ and H atoms are formed; these radicals immediately attack the IL solvent. All these reaction mechanisms contribute to deterioration of neat ILs. Our scenario suggests that this damage may not be fully prevented or mitigated by addition of charge scavengers and excited state quenchers, as such processes as deprotonation are very fast and already compete with very rapid charge transfer. It should be noted however that complete prevention or mitigation is a goal unlikely to be achieved in any system; the real question is whether ionic liquids afford a higher level of radiolytic stability (and separations performance) than TBP/hydrocarbon extraction systems. The studies reported here are effective at quantifying relative yields of early radical radiolysis products, but not as effective, on their own, at establishing the absolute rate of radiolytic damage accumulation in ILs as compared to other systems.

On a more positive note, the damage to the solvent does not necessarily lead to inferior performance of the extraction system if the solvent (i) actively protects the extractant from the damage, diverting the damage towards itself, and (ii) does not yield products that interfere with the metal ion extraction. Our results suggest that this may indeed be the case: rapid, irreversible trapping of the electrons and holes by the solvent means that the extractant is protected. While various radicals are produced, none of these might be expected to react with the alkylphosphate extractant causing dealkylation of the latter. Direct measurements suggest that in 10-40 wt% solutions of alkyl phosphates in these ILs, the yield of the radicals generated via dealkylation of the solute does not exceed a few per cent of the total yield of the (matrix) radicals, implying that the IL solvent takes the brunt of the radiation damage, protecting the solute. This behavior of the IL solvents is fully consistent with our observations:

The DEA involving the alkylphosphates (reaction (8)) is impeded by electron trapping, whereas excitation transfer is impeded by the presence of the cations that have lower excitation energies, serving as quenchers of the excess energy. This quenching may actually protect the extractant even from the *direct* excitation. Thus, the ILs suggest a different paradigm for radiation protection: the complex solvent that actively protects the functional solute (the extractant) in a sacrificial way. The current alkane-phosphate extraction systems involve a highly resistant, structurally simple solvent (the alkane) that efficiently channels *all* radiation damage to the functional solute (the alkyl phosphate). Ionic liquids are the medium in which this pattern is reversed.

## 6. AKNOWLEDGMENTS.

The work at Argonne and Brookhaven is supported by the Office of Science, Division of Chemical Sciences, US-DOE under contracts Nos. DE-AC-02-06CH11357 and DE-AC02-98CH10886, respectively. We thank J. Miller, M. Dietz, R. Chiarizia, and L. Soderholm for many useful discussions and freely given insights, and M. Dietz, P. G. Rickert, A. M. Funston, T. Szreder, M. Thomas, A. Castaño, K. Odynocki, and R. Ramkirath for preparing or providing some of the ionic liquids used in this study. JFW thanks the BNL Office of Educational Programs and the BNL Diversity Office for support of the undergraduate and graduate assistants listed above.

***Supporting Information Available:*** (1) A PDF file containing Figures 1S to 16S with captions and (2) the Appendix. This material is available free of charge via the Internet at http://pubs.acs.org.

**Figure captions.**

**Figure 1.**

Chemical structures of the ions composing the ILs studied.

**Figure 2.**

(a) First derivative EPR spectra (9.445 GHz, modulation 1 G, 0.2 mW) from $C^+$ $NTf_2^-$ solids irradiated by 248 nm laser light at 77K. The spectra were obtained at 70 K. The cations are indicated next to the traces. The spectra are normalized to facilitate the comparison. In the lower trace, the spectrum from $CH_3CH_2CH_2\bullet$ radical observed in irradiated $d_{10}$-pyrene in *n*-butyronitrile at 30 K is shown by a dotted line (from Figure 24S of ref. 49). In the upper two traces, the 67 G doublet of resonance lines from $\bullet NTf_2$ is indicated by open circles (compare with simulations in Figure 2S). The lines from penultimate alkyl radicals in the butyl chain of $P_{14}^+$ are indicated by filled circles (see Figure 5S for simulations). (b) Evolution of the EPR spectrum from $MB_3N^+$ $NTf_2^-$ shown in panel (a) upon warming of the sample to 200 K.

**Figure 3.**

EPR spectra from polycrystalline $Li^+$ $NTf_2^-$ and $Me_3NH^+$ $NTf_2^-$ irradiated by 3 MeV electrons at 77 K (the spectra are obtained at 50 K; 1 G modulation, 0.2 mW). Solid arrows indicate the extreme transitions from $\bullet NTf_2$ radical (see also Figures 1S and 2S). Dashed arrows indicate the sidebands from $^{19}F$ transitions. In both of these spectra, weaker resonance lines from other radicals (see the text) superimpose onto the spectrum of the sulfonamidyl radical.

**Figure 4.**

Formation of trifluoromethyl in low-temperature (77 K) radiolysis (40 s exposure) and photolysis (5 min exposure) of $C^+$ $NTf_2^-$ solids (see section 2 for more detail). The EPR spectra were obtained using 10 G modulation and 20 mW microwave power. The traces were normalized by the maximum signal at the center of the spectrum. The resonance signals from alkyl radicals and the H atoms are removed. (a) In the lower trace, the dashed signal indicates the spectrum and the solid line shows the same spectrum scaled by a factor of 10. The six groups of lines (AA', BB', and CC', panel (b)) from •CF$_3$ are clearly seen. The same lines appear for all ILs including NTf$_2^-$ anion. The solid lines are for electron radiolysis (the scaling factors are 10, 20, 10, and 10, from top to bottom) and the dotted lines are for photolysis (the scaling factors are 40, 80, and 40, from top to bottom). Compare with the simulations shown in Figure 2S for relative intensities. (b) The lines from •CF$_3$ radical in polycrystalline Et$_3$NH$^+$ NTf$_2^-$ at 70 K (solid line) and 130 K (dashed line) irradiated using 3 MeV electrons to the dose of 0.15 Mrad. The signal at the center is removed. The signals inside the circles are (saturated, overmodulated) EPR signals from H atoms. The lines indicated by arrows have not been classified. The AA' doublet corresponds to the extreme transitions occurring when the field is aligned with the symmetry axis.

**Figure 5.**

Evolution of the EPR spectrum from Et$_3$NH$^+$ NTf$_2^-$ shown in Figure 4b upon warming of the sample. The arrows indicate the BB' and CC' lines from trifluoromethyl. At 180 K, many interfering signals disappear and rotation becomes unhindered, so the resonance lines of the alkyl radicals (superimposed onto the singlet) are clearly seen (indicated by open circles).

**Figure 6.**

EPR spectra observed in radiolyzed (a) imidazolium and (b) pyrrolidinium salts. The dashed lines are for 1 G modulation and 0.2 mW, the solid lines are for 10 G modulation and 20 mW; the temperature is 70 K. The lines from trifluoromethyl are indicated by arrows. The question marks indicate unidentified doublet of lines from Figure 3b. The integration of the signals suggests that the yield of •CF$_3$ radicals is 5-10% of the yield of the alkyl and C$_{10}$mim• radicals.

**Figure 7.**

Comparison between EPR spectra obtained in low-temperature photolysis (dashed line) and radiolysis (solid line) of glassy $MB_3N^+$ $NTf_2^-$ at 70 K. The arrows indicate the resonance lines from trifluoromethyl radical. In the radiolyzed solid, there are additional lines (indicated by open circles) from penultimate (ii) C-centered radicals that are missing in the photolyzed solid, where only terminal radicals (i) are observed.

**Figure 8.**

Generation of methyl radicals in low-temperature 3 MeV electron radiolysis (solid line, the upper panel) and 248 nm laser photolysis (solid line, the lower panel) of low-temperature $C_{10}mim^+$ $NTf_2^+$ glass containing 35 wt% trimethylphosphate (TMP). These EPR spectra were observed at 50 K; the irradiation was carried out at 77 K. The positions of the four resonance lines from the methyl radical are indicated by the arrows. The relative yield of these radicals is ca. 2% of the yield of the $C_{10}mim•$ radical. In the lower panel, the dashed line indicates the spectrum of the methyl radical generated by DEA to iodomethane in the same IL glass.

**Figure 9.**

Isodensity surfaces of spin density for $C_s$ symmetrical radical dications of (a) $Et_3NH$, (b) methyltripropylammonium (a model for $MB_3N$), (c) $P_{14}$, and (c) $C_4mim$. The geometries of these species were optimized using B3LYP/6-31+G** method. The isodensity surfaces are for (a) 0.02, (b) 0.01, (c) 0.05, and (d) 0.15 (in atomic units). See the Appendix in the Supplement for atomic charge and spin densities.

Figure 1. Shkrob et al.

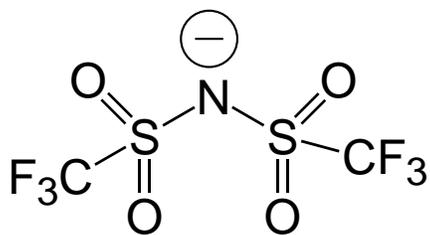

NTf$_2^-$

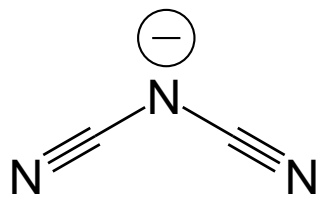

DCA$^-$

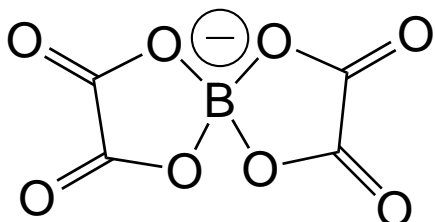

BOB$^-$

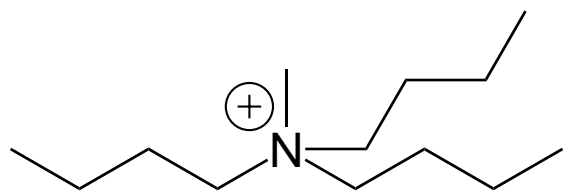

MB$_3$N$^+$

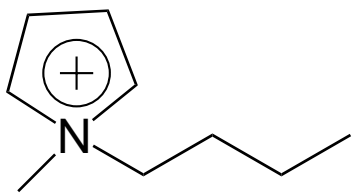

P$_{14}^+$

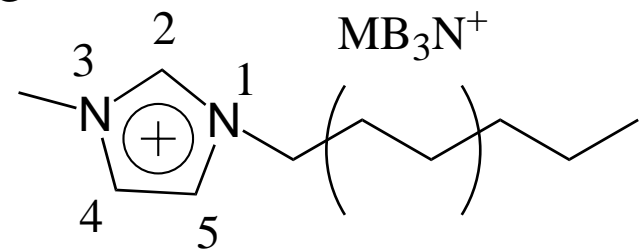

C$_n$mim$^+$

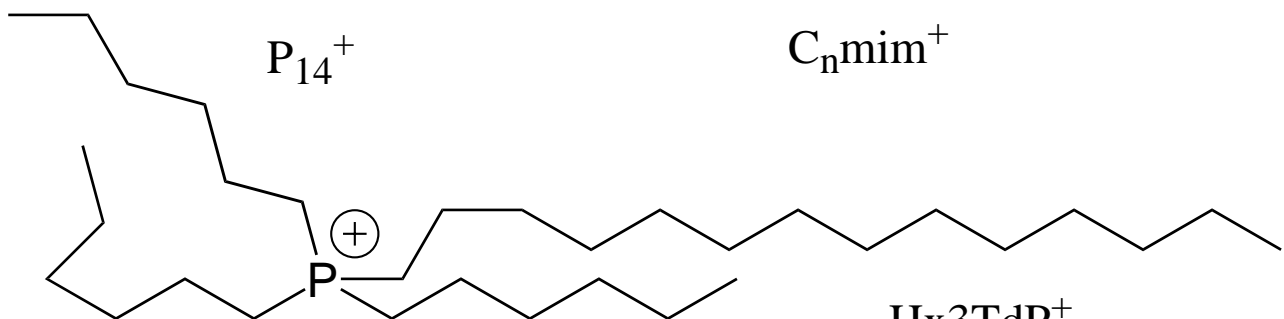

Hx3TdP$^+$

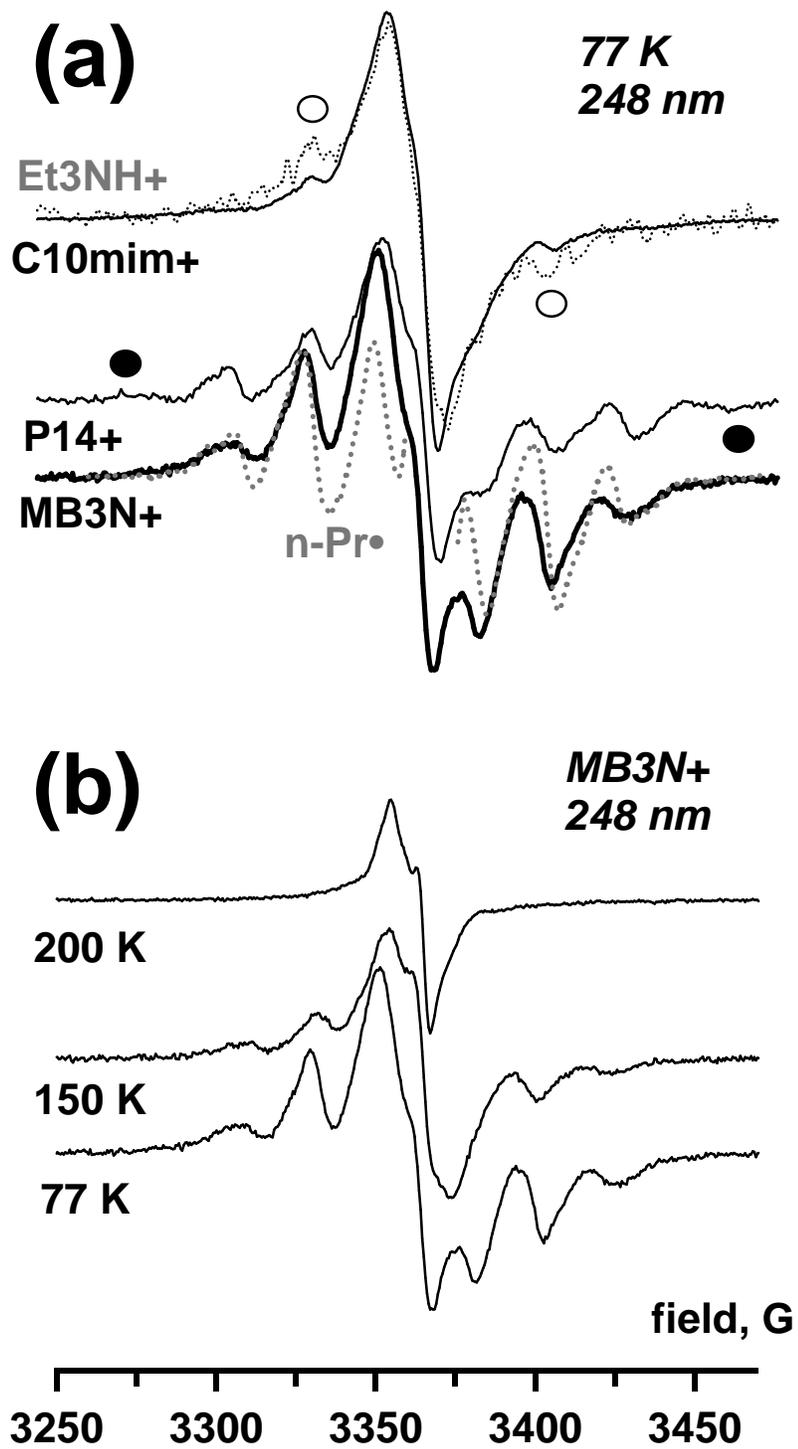

Figure 3. Shkrob et al.

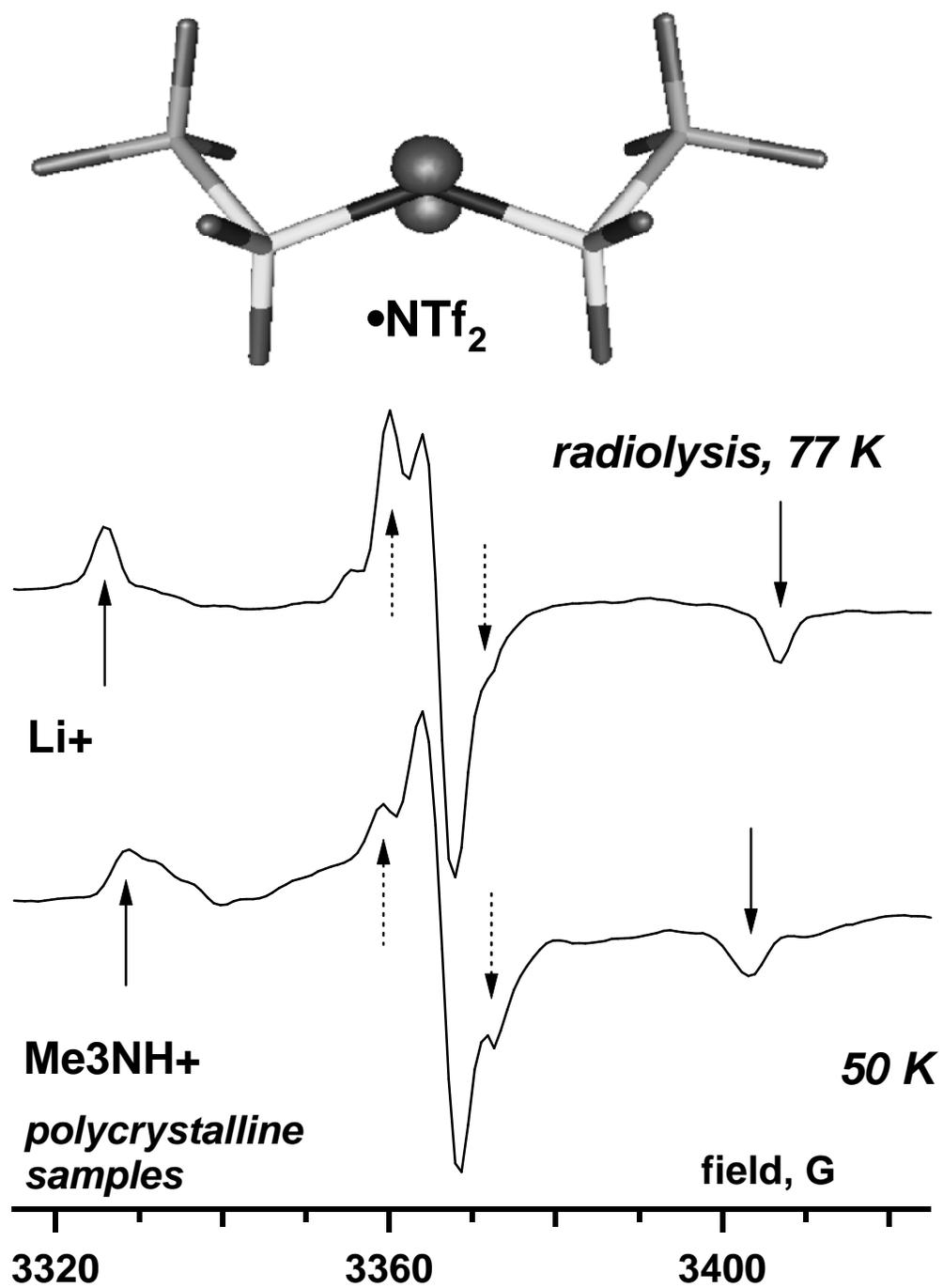

Figure 4. Shkrob et al.

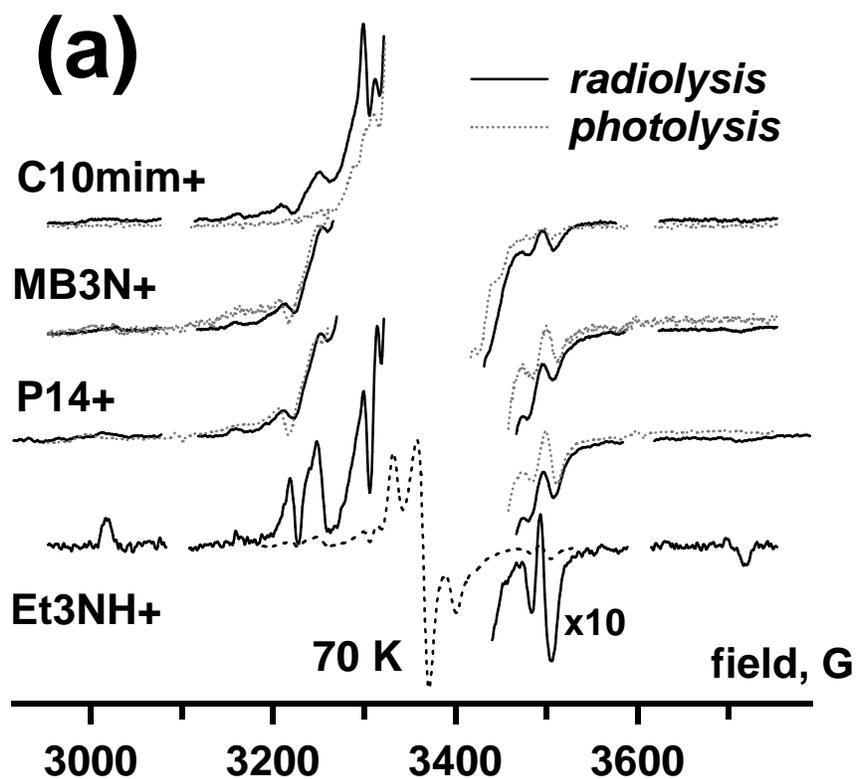

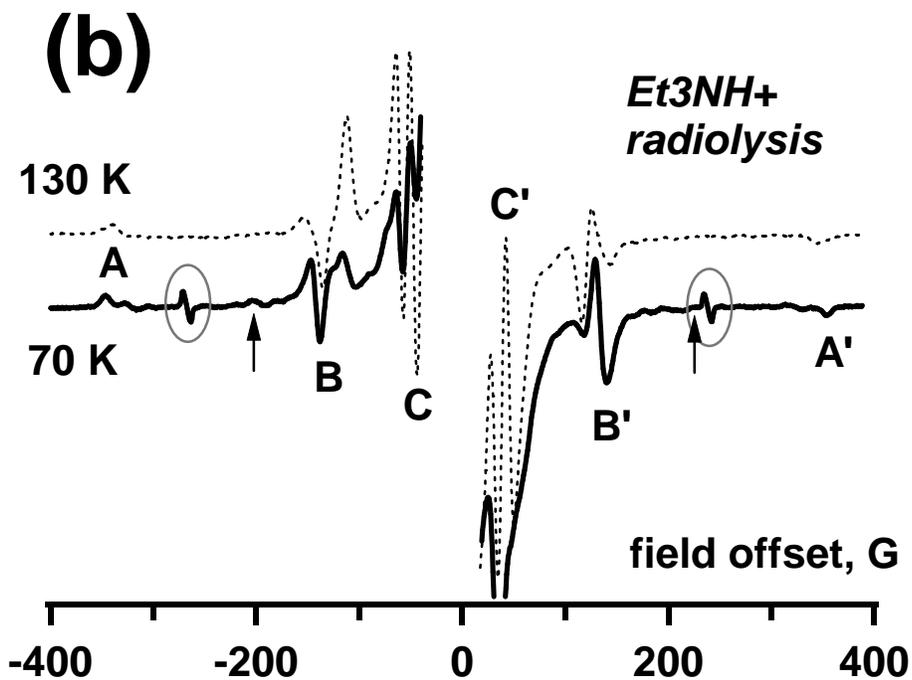

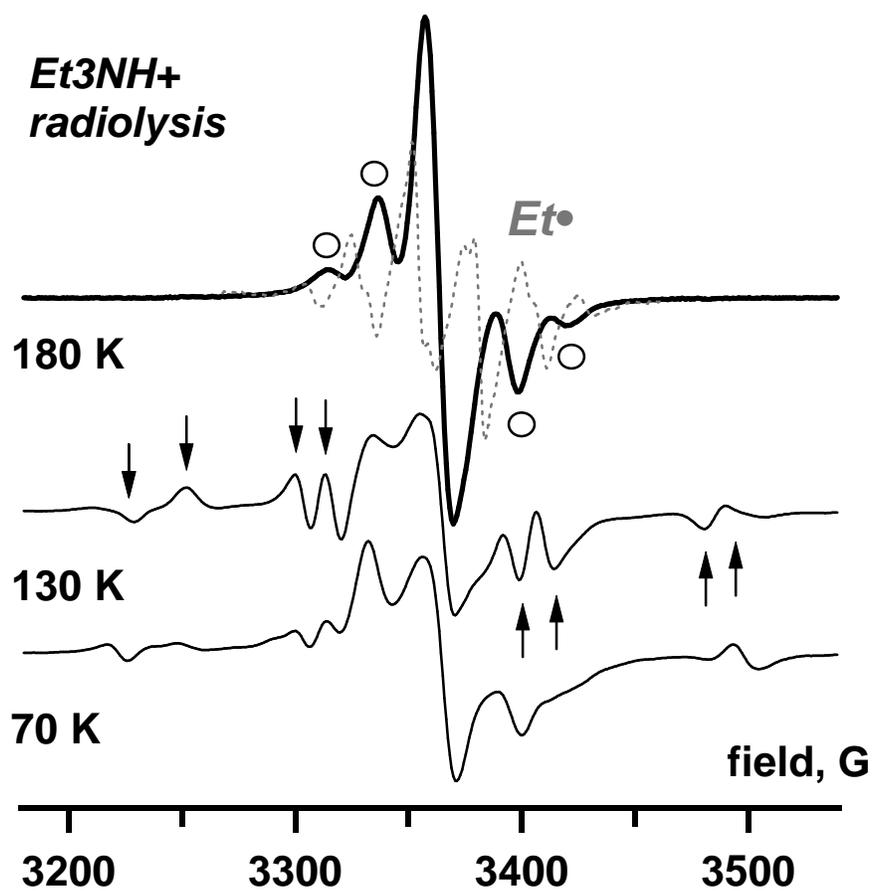

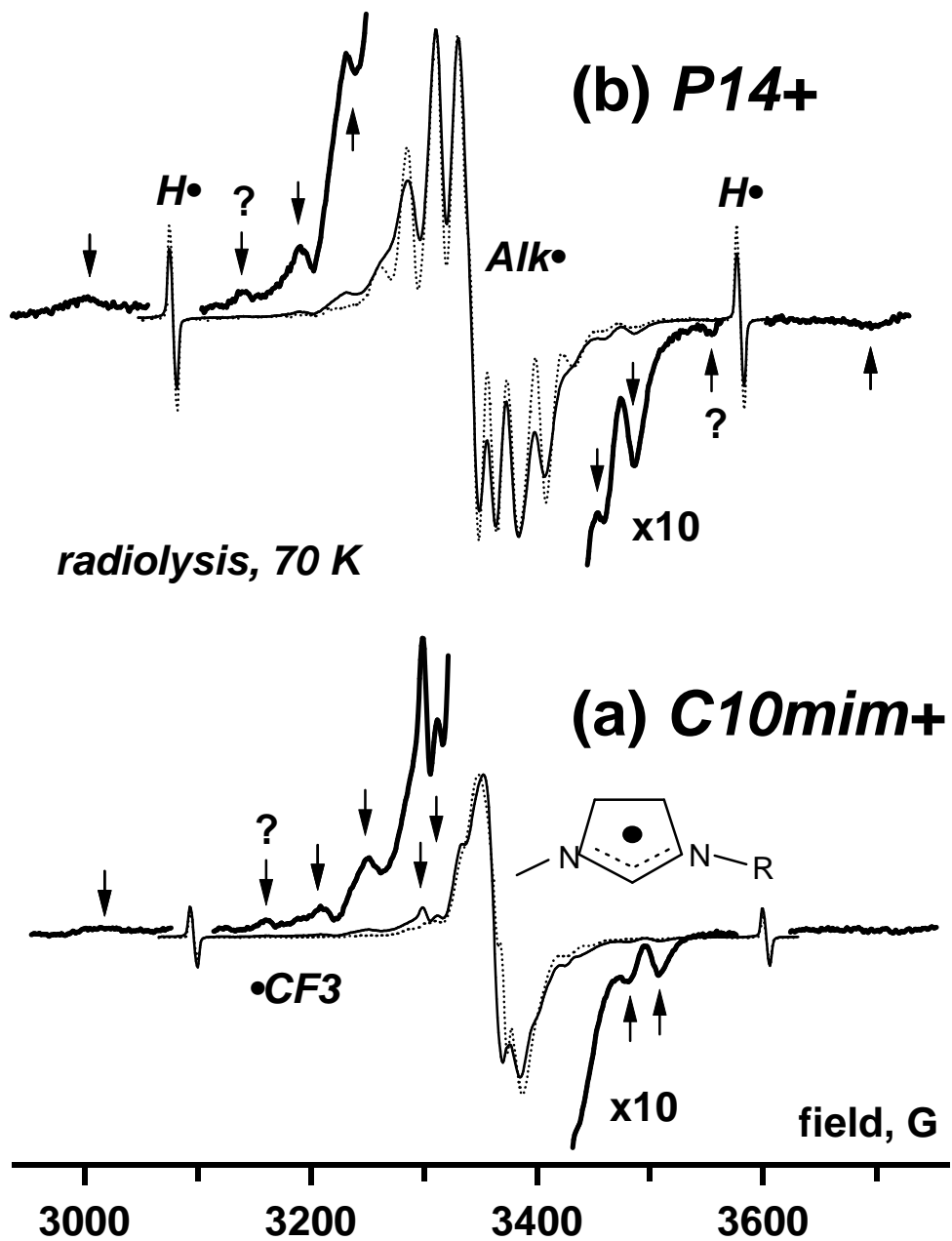

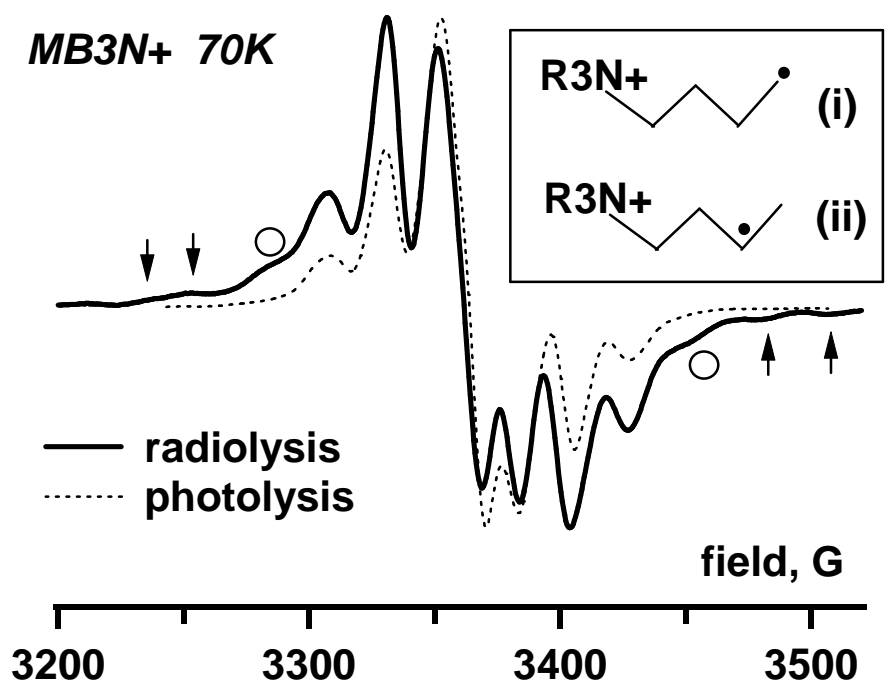

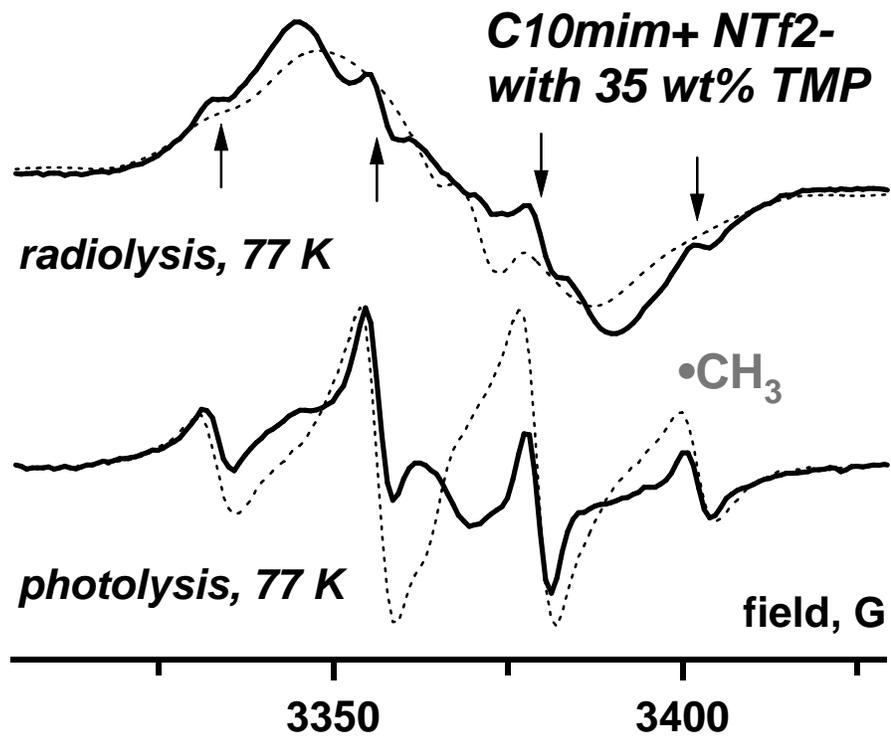

Figure 9. Shkrob et al.

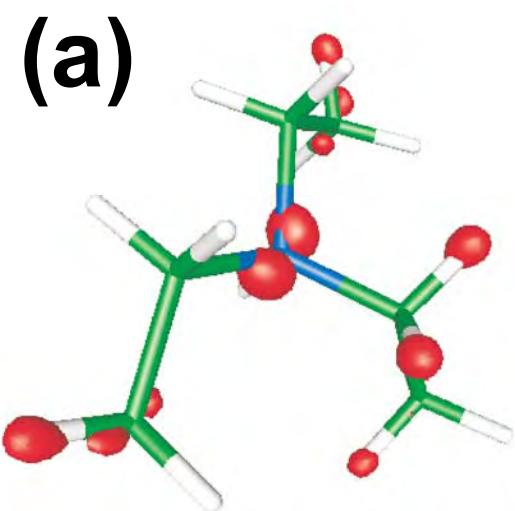
(a)

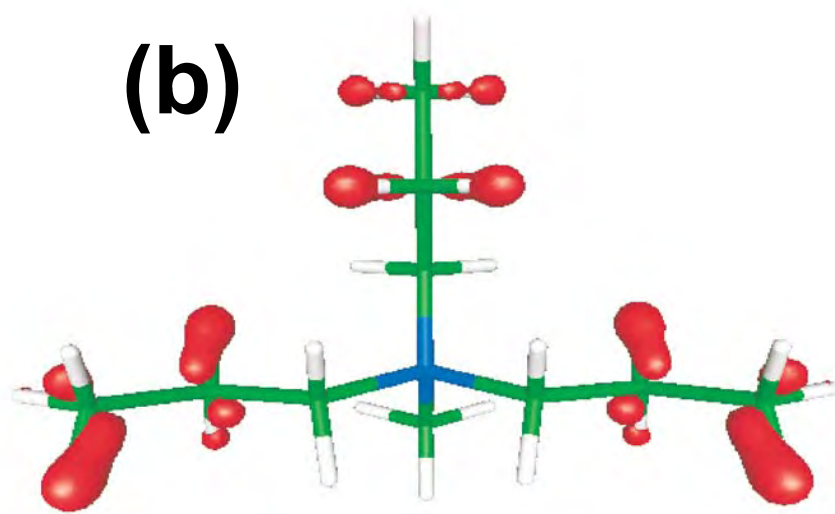
(b)

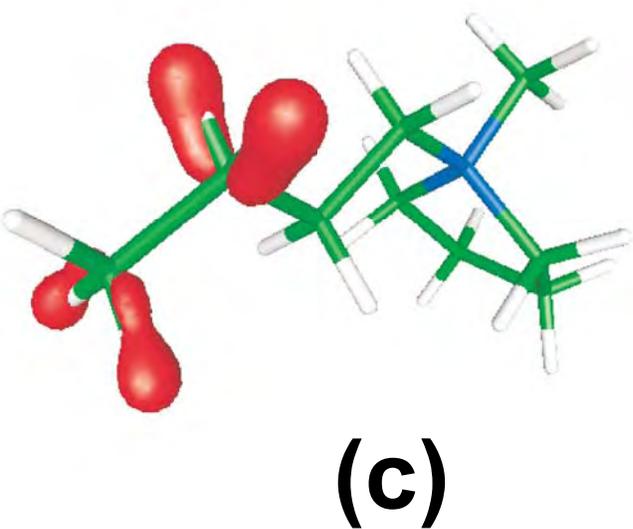
(c)

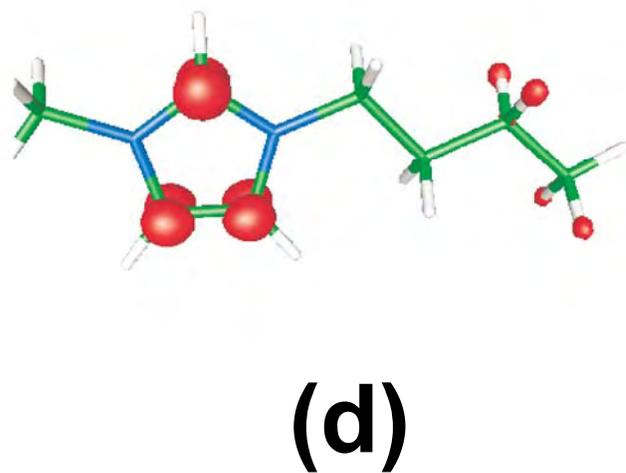
(d)

# SUPPLEMENTARY MATERIAL

**Supporting Information.**

## Figure Captions (Supplement)

**Figure 1S.**

EPR spectra observed in polycrystalline lithium-7 bis(triflyl)amide ($Li^+$ $NTf_2^-$) irradiated by 3 MeV electrons at 77 K and observed at 50, 130, and 165 K, as indicated in the plot. The solid lines are EPR spectra obtained using 1 G modulation and 0.2 mW microwave power; the dashed lines are the same spectra obtained using 10 G modulation and 20 mW microwave power. The side lines (also shown in Figure 3) at the center resonance from the •$NTf_2$ radical become better resolved at the higher temperature, as the trifluoromethyl groups begin to rotate more freely; above 120 K, one can also see a complex pattern of weaker lines from another radical (or a radical pair) with a broad spectrum.

**Figure 2S.**

Simulation of powder EPR spectrum from the $C_{2v}$ symmetrical •$NTf_2$ radical ($b_1$ SOMO) using hfcc parameters estimated using the B3LYP/6-31+G** method. Trace (i) shows the histogram of EPR transitions, and traces (ii) and (iii) show simulated first derivative EPR spectra obtained by convoluting this histogram with Gaussian functions having the width of 1 G and 10 G, respectively. The outer doublet corresponds to the alignment of the magnetic field along the N 2p orbital of the radical which is perpendicular to the SNS plane of the molecule.

**Figure 3S.**

EPR spectra from low-temperature polycrystalline and amorphous trimethylamine bis(triflyl) amide, $Me_3NH^+$ $NTf_2^-$. (a) The comparison between the EPR spectra obtained in the samples following (the top trace) 248 nm laser photolysis of a melt (that was rapidly quenched by immersion into liquid nitrogen) and (b) polycrystalline solid irradiated by 3 MeV electrons (at 77 K). These EPR spectra were observed at 50 K (1 G modulation and 0.2 mW microwave power). In the lower trace, the lines from the trifluoromethyl radical are scaled by a factor of 10. The narrow resonance lines corresponding to the extreme transitions in the •$NTf_2$ radical (Figure 2S) are clearly seen in the polycrystalline (powder) samples, but these lines are poorly resolved in the vitreous sample. (b) Evolution of the EPR spectrum from 3 MeV electron irradiated $Me_3NH^+$ $NTf_2^-$ upon warming the sample from 50 K to 150 K to 300 K (the upper EPR spectrum was obtained at 200 K after warming the sample to 300 K for 5 min). The arrows indicate the lines from the •$CF_3$ radical, the filled circles indicate the lines from the •$NTf_2$ radical. The open circles indicate the lines from •$NMe_3^+$ or •$NHMe_2^+$ radical cations which are observed above 120 K.

**Figure 4S.**

Simulation of the first derivative EPR spectra form the terminal and penultimate alkyl radicals and the •$CF_3$ radical. These simulated spectra were normalized by the integral of the EPR signal, i.e. their relative amplitudes correspond to a 1:1:1 mixture of these three radicals (the same inhomogeneous broadening of 10 G was assumed for all resonance lines). The powder spectrum of trifluoromethyl was simulated using the hfcc parameters obtained from DFT calculation in the Appendix. The spectra of alkyl radicals were simulated assuming complete averaging of their anisotropic tensors; the hf constants were taken from reference 41. For terminal radical, we let $a(2H_\alpha)$=22 G, $a(2H_\beta)$=29.1 G, and $a(2H_\gamma)$=0.7 G, for penultimate radical, we let $a(H_\alpha)$=21.9 G, $a(3H_\beta)$=24.6 G, $a(2H_\beta)$=27.2 G, and $a(3H_\gamma)$=3.9 G. The AA', BB', and CC' doublets in the spectrum of trifluoromethyl are indicated. Isotropic $g$-tensors were assumed for all of these radicals.

**Figure 5S.**

(a) Ethyl, (b) methyl, and (c) methyl-$d_3$ radicals generated in frozen ILs. All of these EPR spectra were observed at 50 K for field modulation of 1 G, and the microwave power of 0.2 mW. These radicals were generated via DEA to the halocarbon dopants. Normalized EPR spectra from 3 MeV electron irradiated (a) $C_{10}mim^+$ $NTf_2^-$ (red dashed line) and $P_{14}^+$ $NTf_2^-$ (solid black line) containing 10 wt% EtBr and (b) $P_{14}^+$ $NTf_2^-$ (thin blue line) containing 10 wt% $CH_3I$ are shown. The four resonance lines (1:4:4:1) from the methyl radical (marked by the arrows) are superimposed onto the spectrum of alkyl radicals that are derived from $P_{14}^+$. The latter signals can be removed by subtracting the scaled EPR spectrum obtained in irradiated neat IL to produce the EPR spectrum shown by the bold (black) line which is mainly from the methyl radical. To demonstrate that the EPR spectrum from the doped sample shown in (b) is composite (as there is a possibility that the methyl radical abstracts hydrogen from the cation), in (c) we show the seven line EPR spectrum from the •$CD_3$ radical obtained by 248 nm laser photolysis [49] of 14 wt% $CD_3CN$ in $P_{14}^+$ $NTf_2^-$. The resonance lines from the trideuteromethyl radical (with a peak-to-peak separation of 3.3 G vs. the literature value of 3.4 G) are superimposed on the spectrum from (mostly) •$NTf_2$. The stability of the methyl radical in the IL matrices is thus demonstrated.

**Figure 6S.**

(a) Simulation of EPR spectra from the radicals derived from the aliphatic chains of the amines/alkylammonium using the hfcc's reported in the literature (ref. 41, vol. II/9b, section 3 and ref. 49). Only isotropic hfcc's were taken into account. In panel (b), we compare the spectra from $Et_2NH^+CH•CH_3$ and $Et_2NH^+CH_2CH_2•$ radicals calculated using the hfcc's obtained in our DFT calculations (see the Appendix). The parameters for the terminal radical are indicated in the plot.

**Figure 7S.**

Isodensity SOMO maps for (a) $C_4mim•$ radical and (b,c) two -H radicals obtained via the loss of imidazolium protons. See Figure 7S for their simulated EPR spectra.

**Figure 8S.**

(a) Simulated powder EPR spectra for imidazolium radicals. The lower trace is for $C_4mim\bullet$ radical; the upper three traces are for -H radicals, as indicated in the figure (see also Figure 7S). The spectra were simulated assuming 1 G broadening. In panel (b), the same spectra are shown assuming 10 G broadening. Trace (i) is for the imidazolium radical; traces (ii) are -H radicals.

**Figure 9S.**

EPR spectrum of polycrystalline $Bu_4N^+$ $NTf_2^-$ irradiated by 3 MeV electrons at 77 K (observed at 50 K, for the field modulation of 1 G and the microwave power of 0.2 mW). The sharp "spikes" are resonance lines from H atoms in the sample tube. The arrows indicate the lines from $\bullet CF_3$. The dashed (blue) line is the EPR spectrum from $MeBu_3N^+$ $NTf_2^-$ glass. Despite the difference in the sample morphology and the substitution at the nitrogen, the spectra of the resulting alkyl radicals are very similar.

**Figure 10S.**

The comparison of EPR spectra from radiolyzed $MB_3N^+$ $NTf_2^-$ and $P_{14}^+$ $NTf_2^-$ glasses (70 K; the field modulation and the microwave power are indicated in the plot). The arrows indicate the resonance lines from trifluoromethyl. The filled circles indicate the outer lines from the penultimate C-centered radicals in the *n*-butyl groups of these cations.

**Figure 11S.**

(a) A comparison between the EPR spectra obtained from 3 MeV electron irradiated polycrystalline sodium dicyanamide (DCA) and 248 nm laser irradiated IL glass composed of $P_{14}^+$ $DCA^-$. In the upper trace, spin transitions assigned to the $\bullet N(NC)_2$ radical are indicated by the arrows. In the lower trace, EPR spectrum observed in laser irradiated $P_{14}^+$ $NTf_2^-$ (dashed line) and radiolyzed $P_{14}^+$ $DCA^-$ (blue line) are shown for comparison. These EPR spectra were obtained at 50 K (field modulation of 1 G and microwave power of 0.2 mW). (b) Simulated EPR spectra of $\bullet N(NC)_2$ (the bold line is for 10 G width, the thin line is for 1 G width) obtained using the hfc tensors calculated by B3LYP/6-31+G** method (see the Appendix). The $b_1$ SOMO of the radical is shown in the inset.

**Figure 12S.**

(a) EPR spectra observed in radiolyzed (solid black line) and laser photolyzed (dashed blue line) frozen $P_{14}^+$ $BOB^-$ glass at 77 K (these EPR spectra were obtained at 50 K, modulation 1 G, and 0.2 mW). The spectrum is dominated by oxygen-centered radicals; observe the difference between the two traces, indicating the composite character of the EPR spectrum. Panel (b) shows the same EPR spectrum scaled by a factor of 25, so that the resonance lines from the alkyl radical are seen in more detail. The yellow line that is superimposed on these spectra is a scaled EPR spectrum obtained from laser-irradiated $P_{14}^+$ $NTf_2^-$ glass at 77 K. It is seen that the same alkyl radicals derived from the $P_{14}^+$ cation moiety are generated in all three of these samples. Integration of the EPR lines indicates that the yield ratio of the alkyl to the O-centered radical(s) is ca. 1:2.3.

**Figure 13S.**

EPR spectra observed in irradiated frozen IL glasses consisting of $Hx_3TdP^+$ cation, trihexyltetradecylphosphonium (50 K, 1 G, 0.2 mW). (a) The alkyl radical generated in radiolyzed (solid black line) and photolyzed (dashed blue line) $Hx_3TdP^+$ $NTf_2^-$ glass. Observe the near absence of EPR signals from the •$CF_3$. (b) Solid lines: EPR spectra from radiolyzed and photolyzed $Hx_3TdP^+$ $BOB^-$ glass (the dashed lines are scaled EPR spectra from alkyl radicals shown in panel (a), respectively). The relative yields of the alkyl and O-centered radicals, as estimated by integration of the EPR spectra, are 1:1.6 (radiolysis) and 1:1.1 (photolysis), respectively.

**Figure 14S.**

Dealkylation of triethyl phosphate (TEP) in irradiated IL glasses at 77 K (the EPR spectra are observed at 50 K). The dashed red lines in both panels show the EPR spectrum of the ethyl radical (generated by 3 MeV electron irradiation of EtBr in $C_{10}mim^+$ $NTf_2^-$ glass, Figure 5S(a)). (a) Laser photolysis (248 nm) and 3 MeV electron radiolysis of $C_{10}mim^+$ $NTf_2^-$ glass containing 10 wt% (blue lines) and 35 wt% (black lines) TEP. While the resonance lines from the ethyl (indicated by the arrows) can be readily observed in the photolyzed samples (due to the direct photoexcitation of the triethyl phosphate), almost no such features are observed in the radiolyzed samples; the relative yield of this radical is < 2% of the yield of the $C_{10}mim$• radical. (b) Radiolysis (solid black lines) and laser photolysis (dashed blue lines) of $P_{14}^+$ $BOB^+$ containing 20 wt% TEP. The microwave power is indicated in the plot (the modulation is 1 G). The open circles indicate the features from the ethyl radical derived from TEP. The relative yield of this radical is < 5% of the total yield of all radicals observed.

**Figure 15S.**

(a) *Upper traces:* EPR spectra of 3 MeV electron irradiated $P_{14}^+$ $NTf_2^-$ glass containing trimethylphosphate (TMP). The solid (black) line is from a sample containing 40 wt% TMP, the dashed (blue) line is from a sample containing 14 wt% TMP; the yellow line is the spectrum for neat IL. The four resonance lines from the methyl radical derived from TMP stand out against the stronger background signal from the (mostly) alkyl radicals derived from the $P_{14}^+$ cation. The bottom two traces are EPR spectra obtained in photolyzed glass containing 40 wt% TMP (solid line) and radiolyzed glass containing 10 wt% $CH_3I$ (dashed line). These two spectra are very similar, as both of the spectra are dominated by the methyl radical that is generated via DEA to the dopant. In panel (b) the upper trace is from $P_{14}^+$ $NTf_2^-$ glass containing 40 wt% TMP irradiated to the radiolytic doses of 0.15 Mrad (solid black line) and 1.25 Mrad (dashed blue line); these EPR spectra were normalized so that the outer lines from the alkyl radicals had the same amplitudes. As the relative yield of the methyl radical among the radical products of radiolysis decreases with the increasing dose, the subtraction of the two spectra yields the EPR spectrum from the (predominantly) methyl radical (trace (ii)). The arrows in the lower trace indicate the four resonance lines from the •$CH_3$. Integration of these EPR spectra indicates that the yield of the methyl radical relative to the alkyl radical (for 0.15 Mrad dose) for 3 MeV electron radiolysis at 77 K is ca. 1% for 14 wt% TMP and 4% for 40 wt% TMP. Though the dealkylation of TMP occurs in radiolysis, most of the damage is directed towards the IL matrix. The EPR spectra were obtained at 50 K, for field modulation of 1 G, and the microwave power of 0.2 mW.

**Figure 16S.**

EPR spectrum of electron beam irradiated $P_{14}^+$ BOB$^-$ glass containing 20 wt% trimethylphosphate (TMP). The spectra were obtained at 50 K, for field modulation of 1 G, and the microwave power of 0.2 mW. In panel (a), the spectrum (solid black line, trace (i)) is shown over the spectrum observed from irradiated neat IL glass (dashed blue line, trace (iii) in both of the panels). The four resonance lines from the methyl radical derived from TMP are indicated by the arrows. In panel (b) the wings of these spectra (scaled by a factor of 10) are shown separately, with EPR spectra from the alkyl radical derived from deprotonated $P_{14}^+$ cation and the methyl radical in $P_{14}^+$ $NTf_2^-$ glass (trace (iv) from Figure 5S, trace (c)) superimposed. Subtraction of this underlying EPR signal from the alkyl radical yields the outer two lines of the methyl radical. Integration of these lines indicates that the relative yields of the alkyl, the O-centered, and the methyl radical in the radiolyzed sample are 1:1.6:0.08. Though the dealkylation of TMP occurs in radiolysis, most of the damage is directed towards the IL matrix, with the methyl radicals accounting for only 3% of the radical products. These EPR spectra were obtained at 50 K, for field modulation of 1 G, and the microwave power of 0.2 mW.

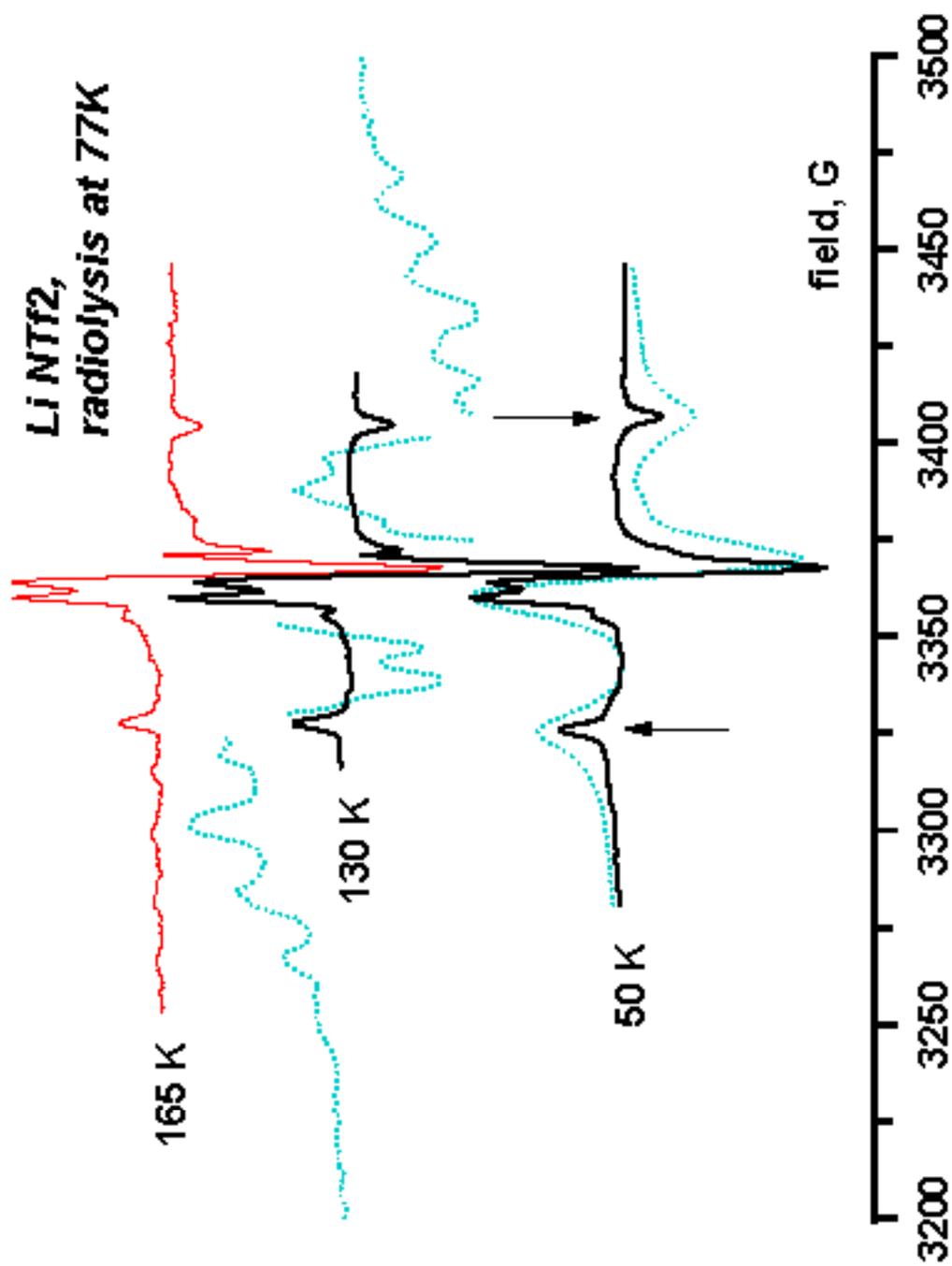

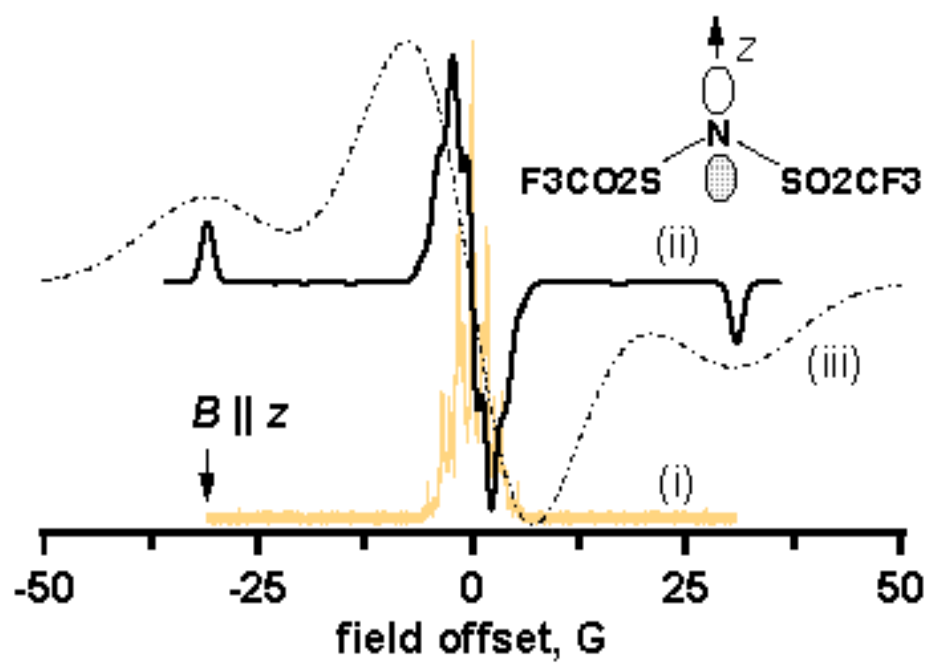

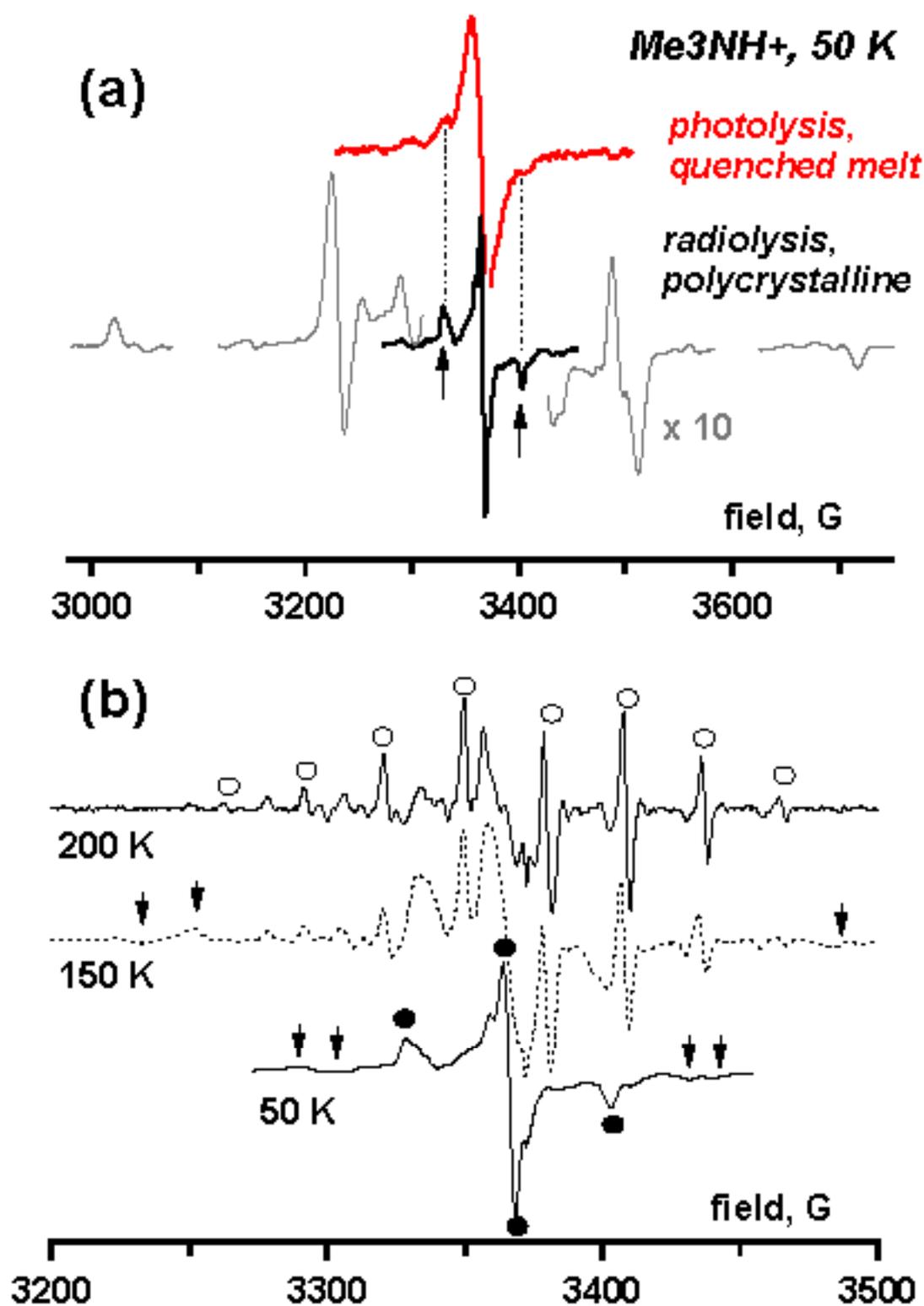

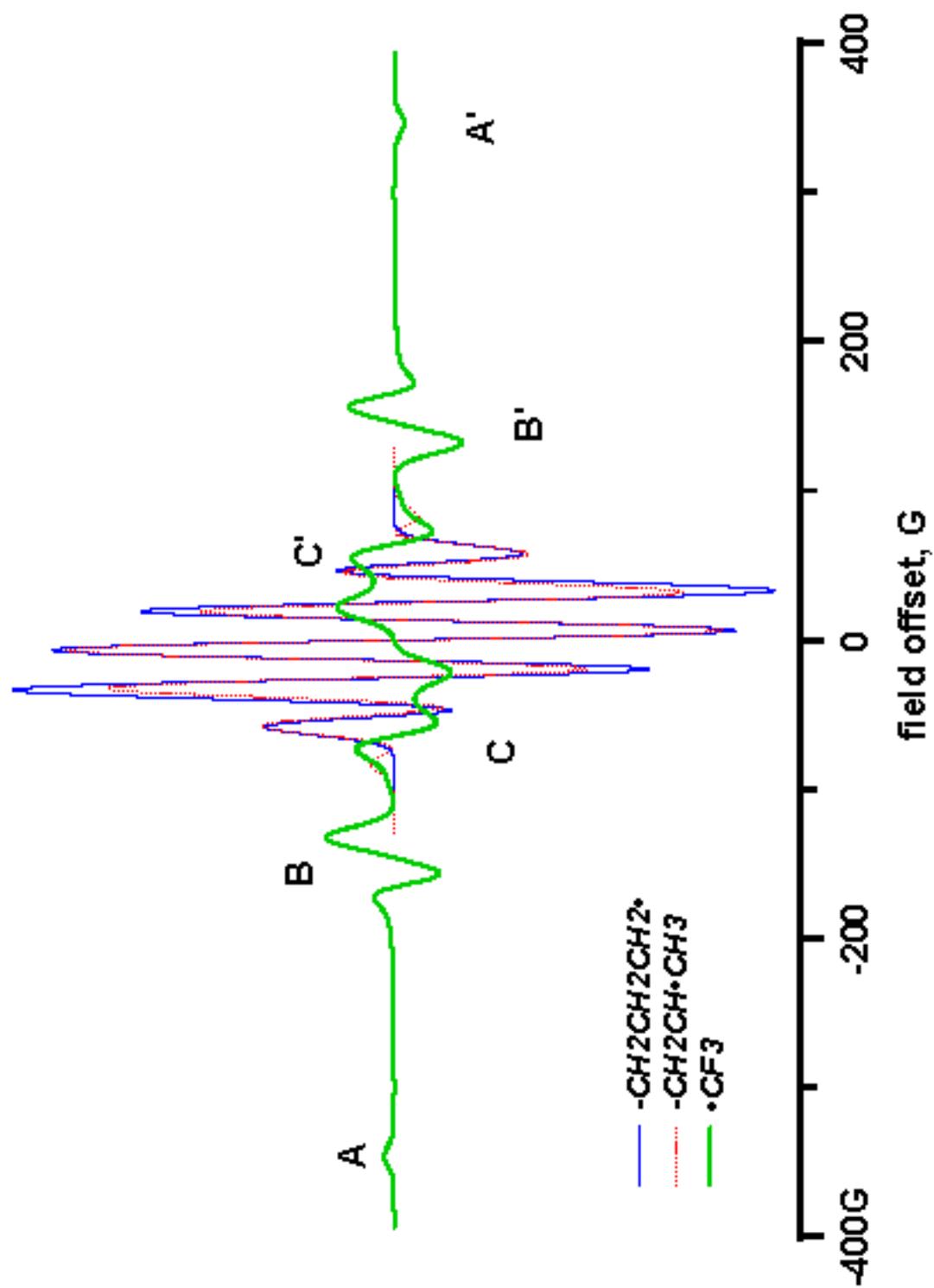

Figure 4S. Shkrob et al.

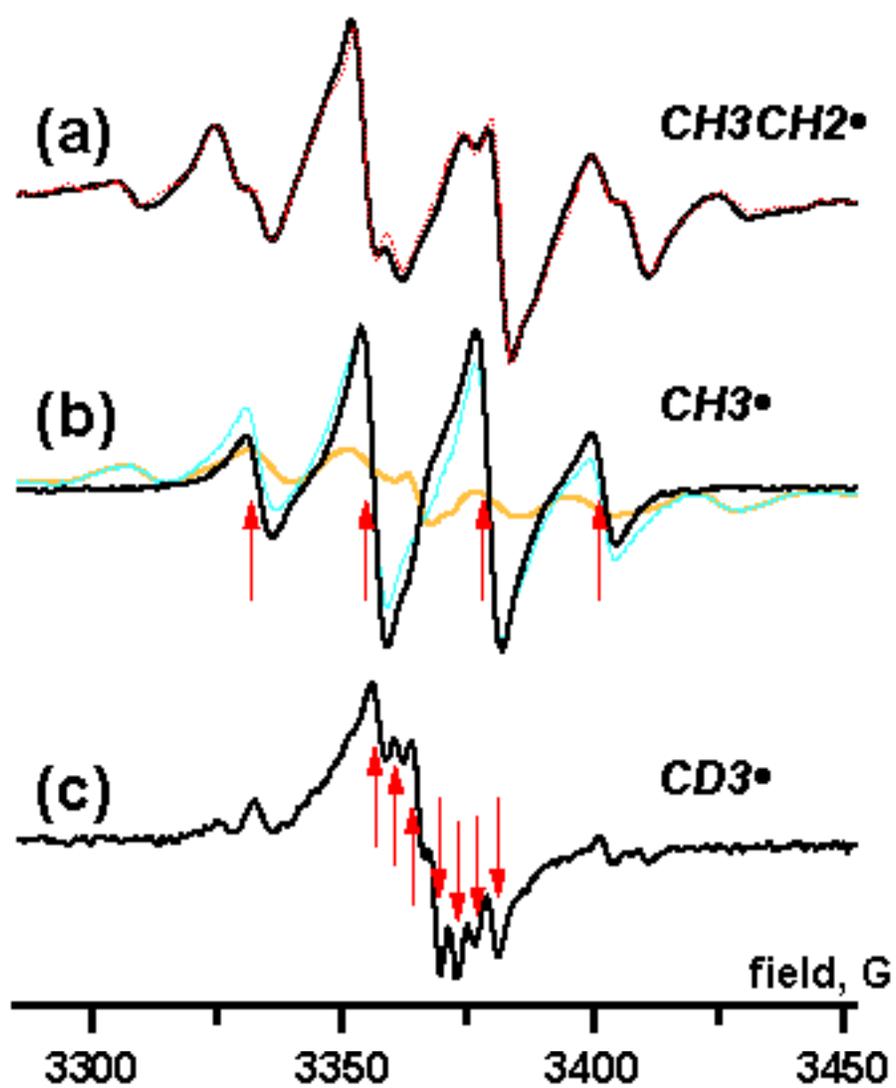

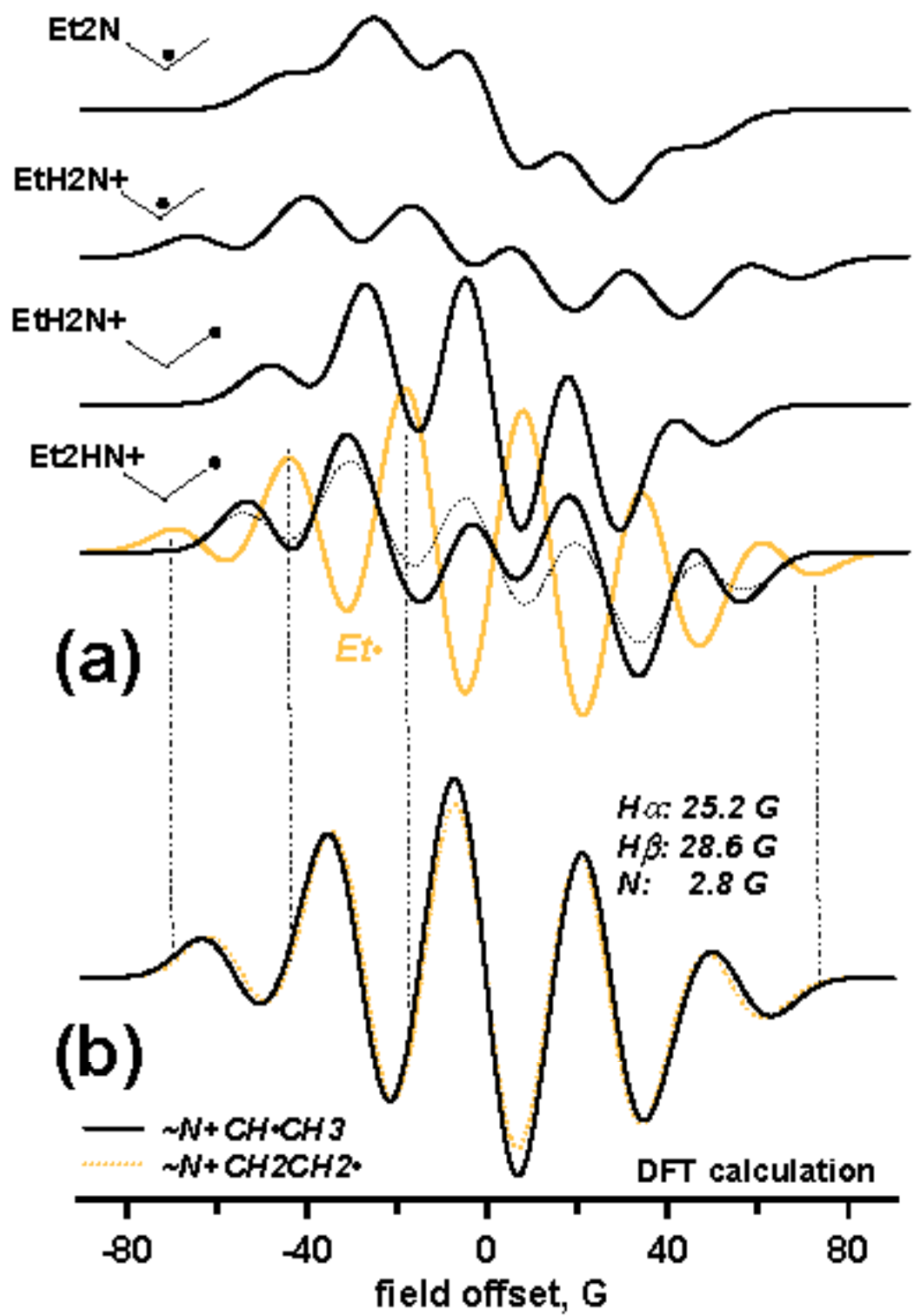

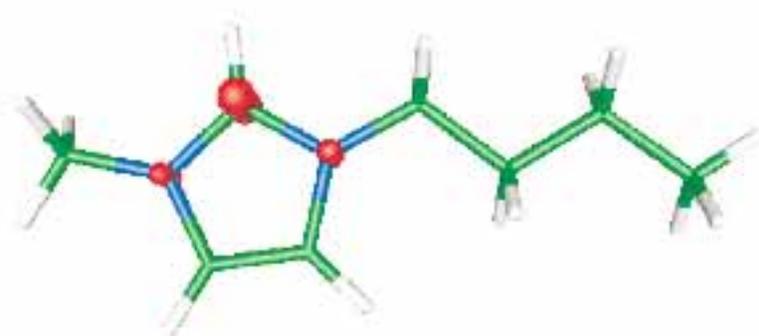 (a)

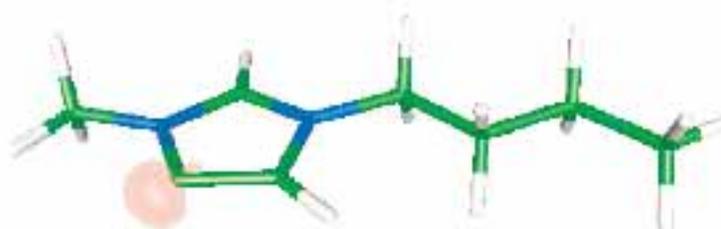 (b)

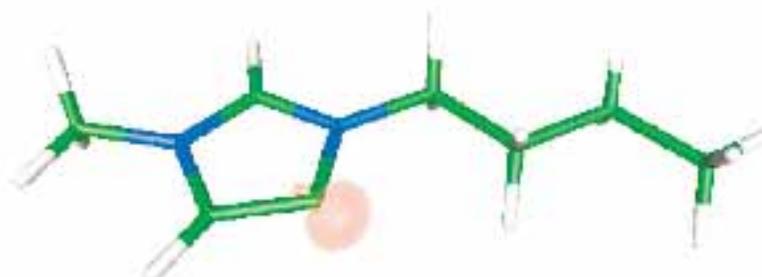 (c)

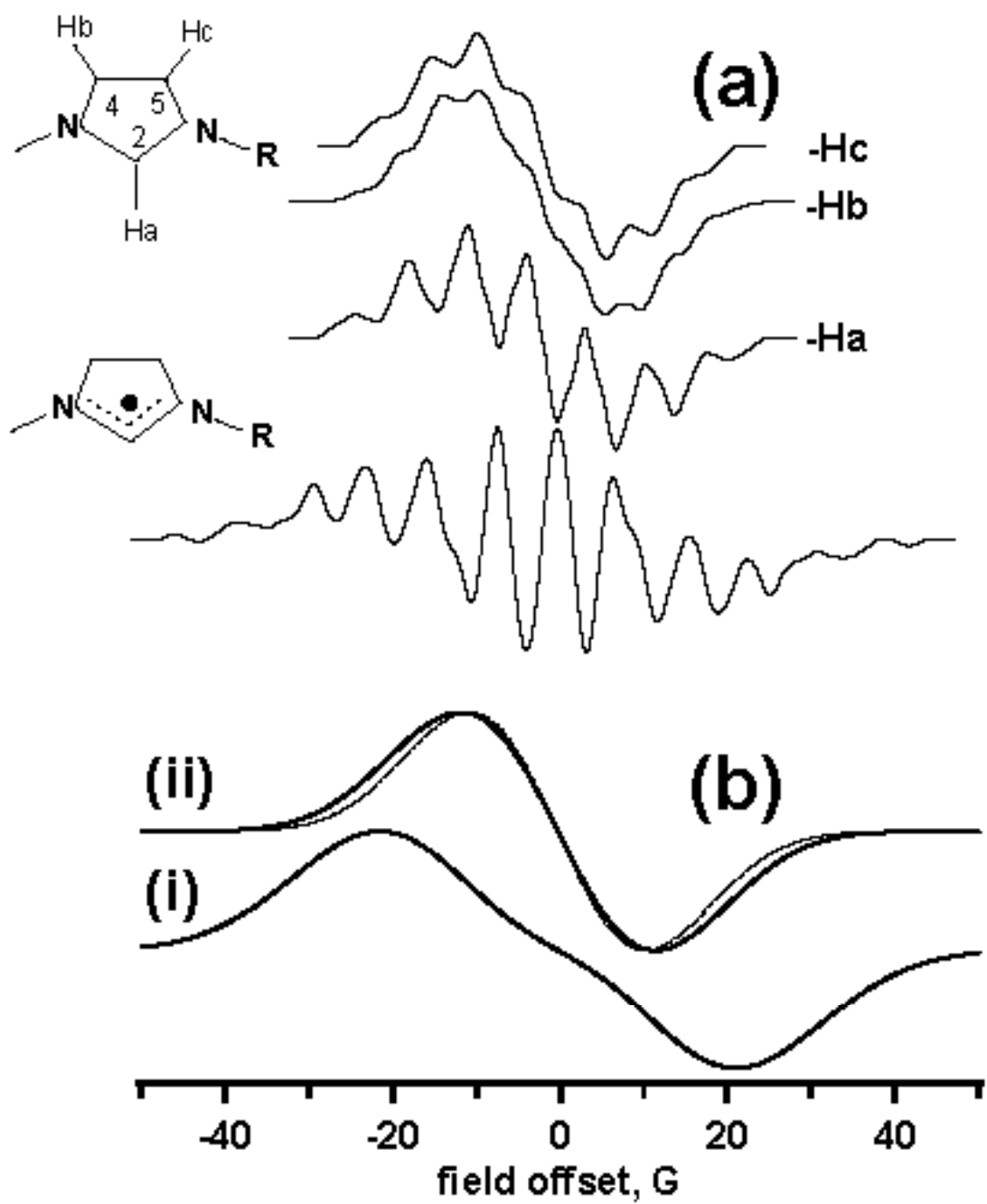

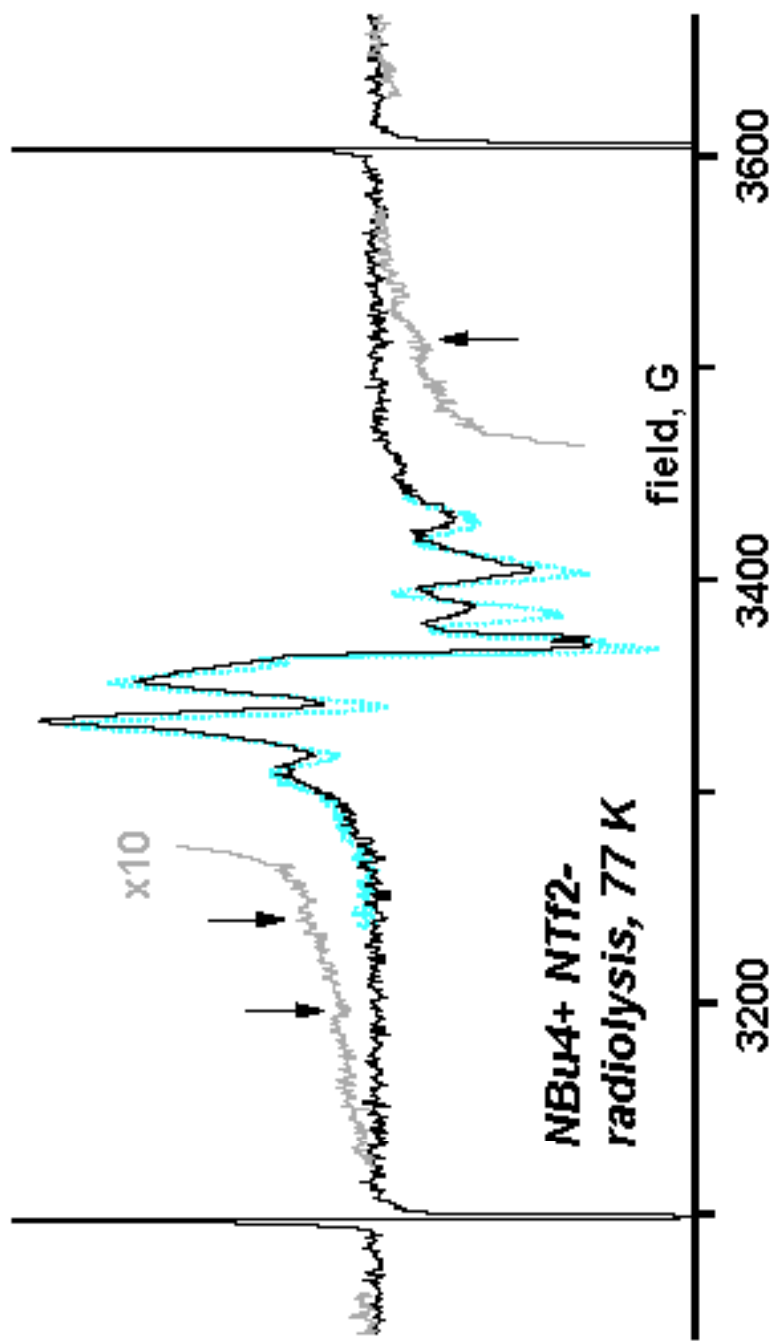

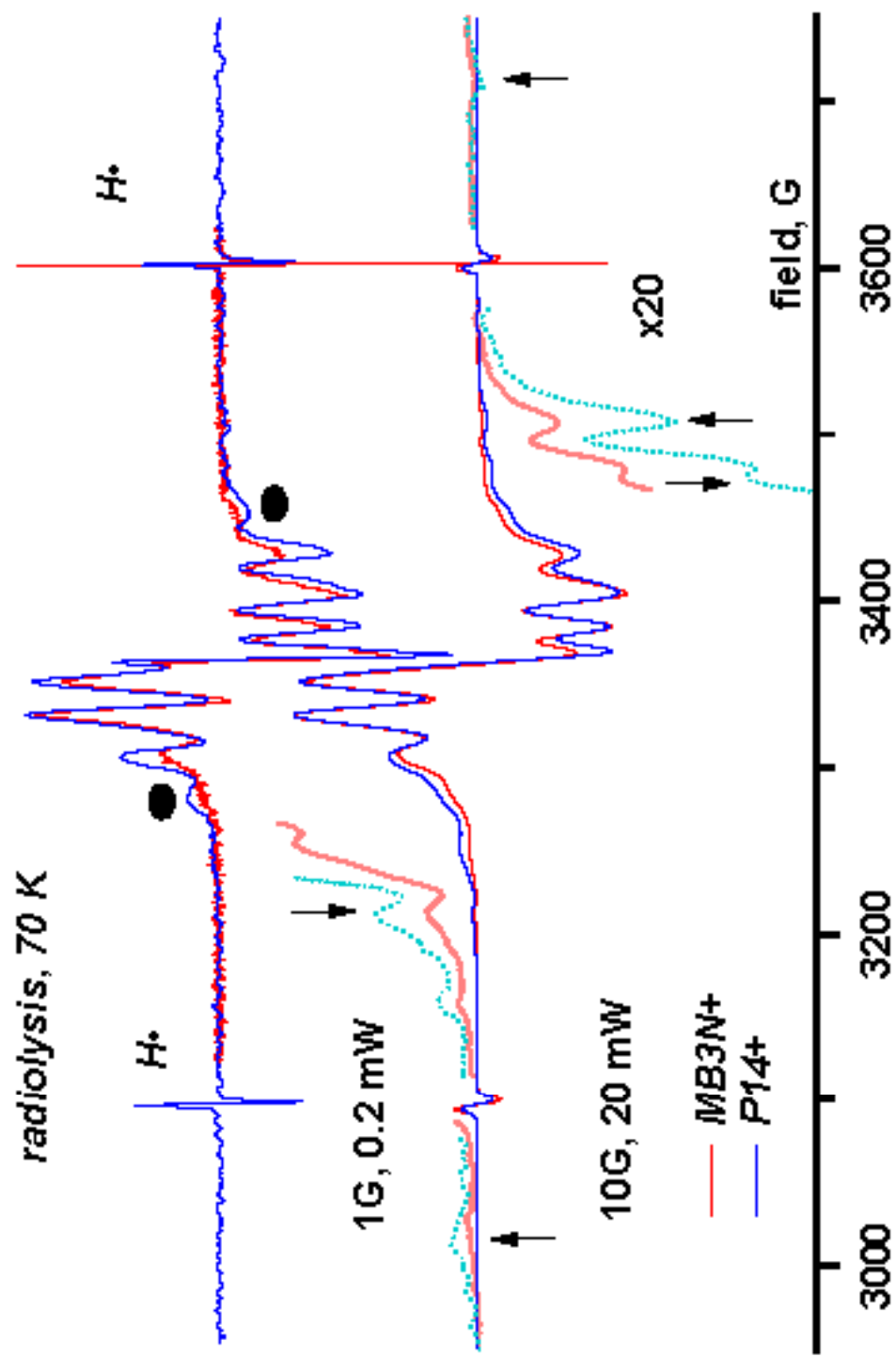

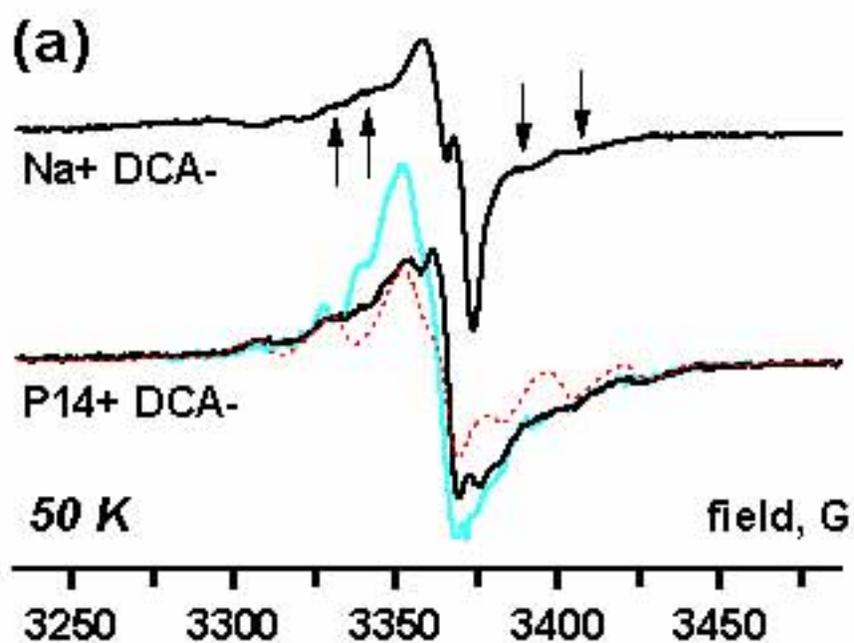
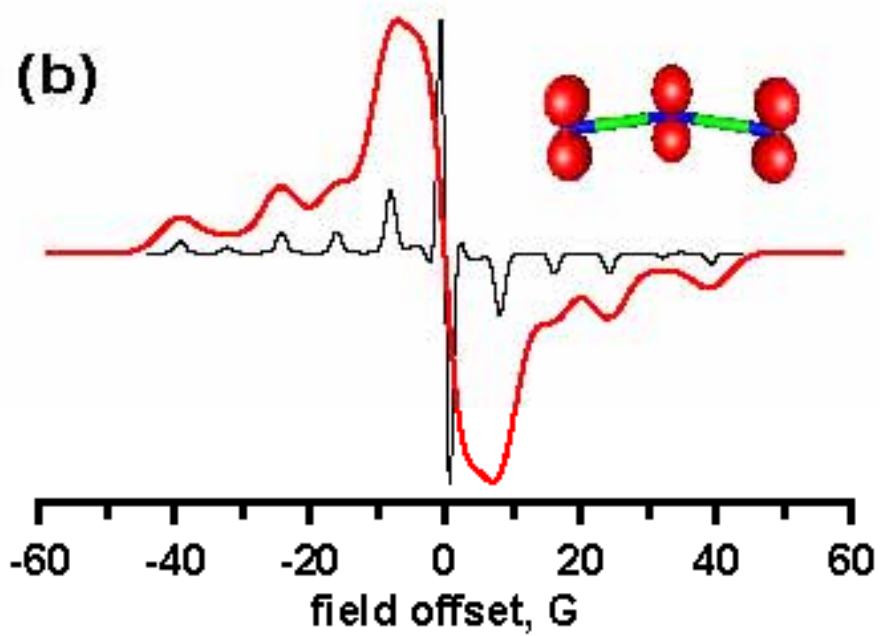

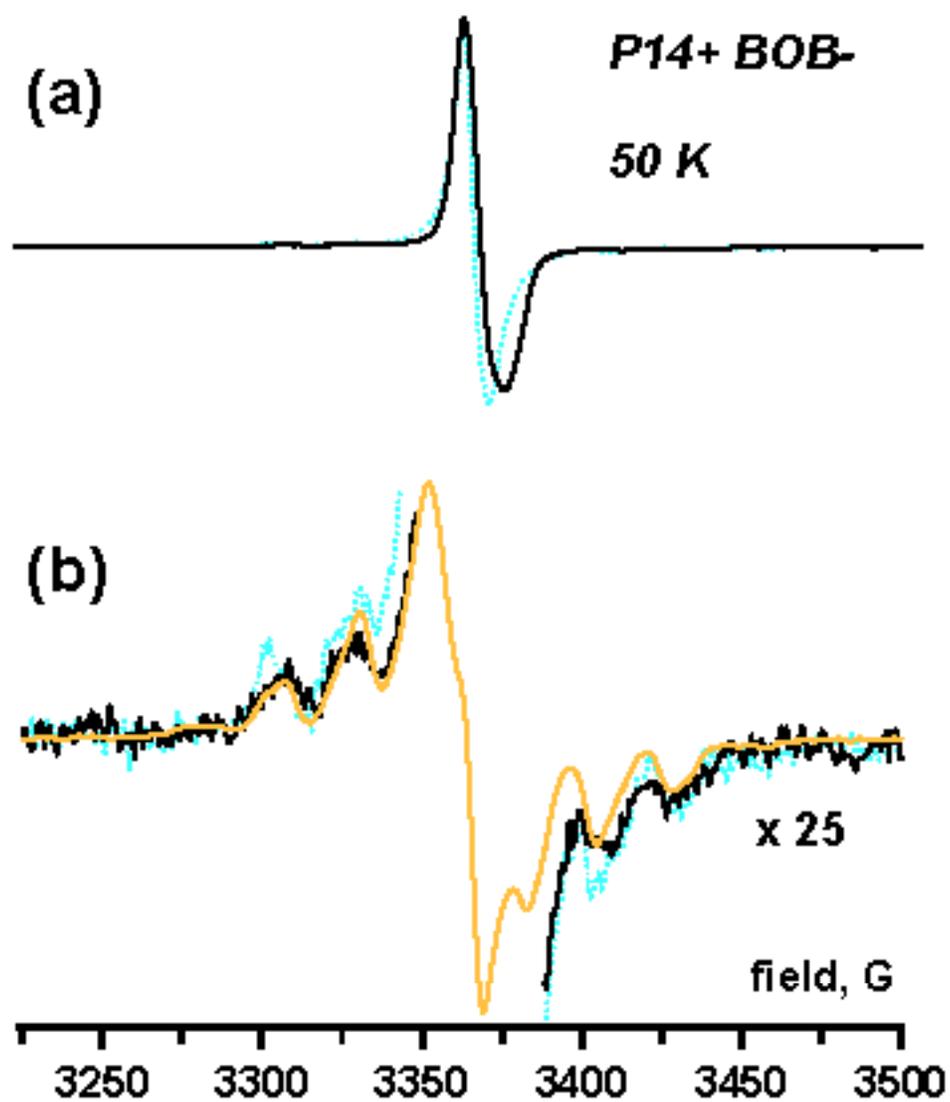

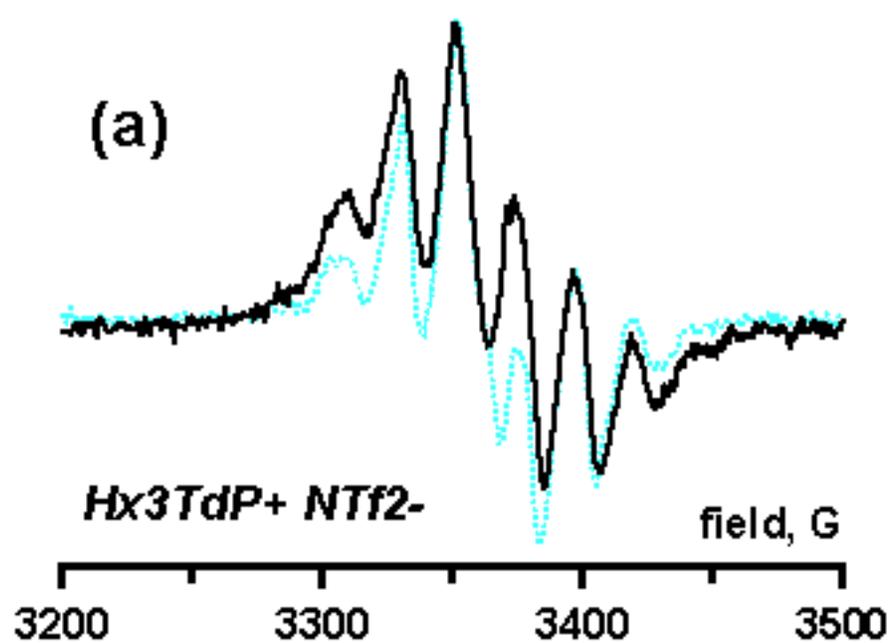

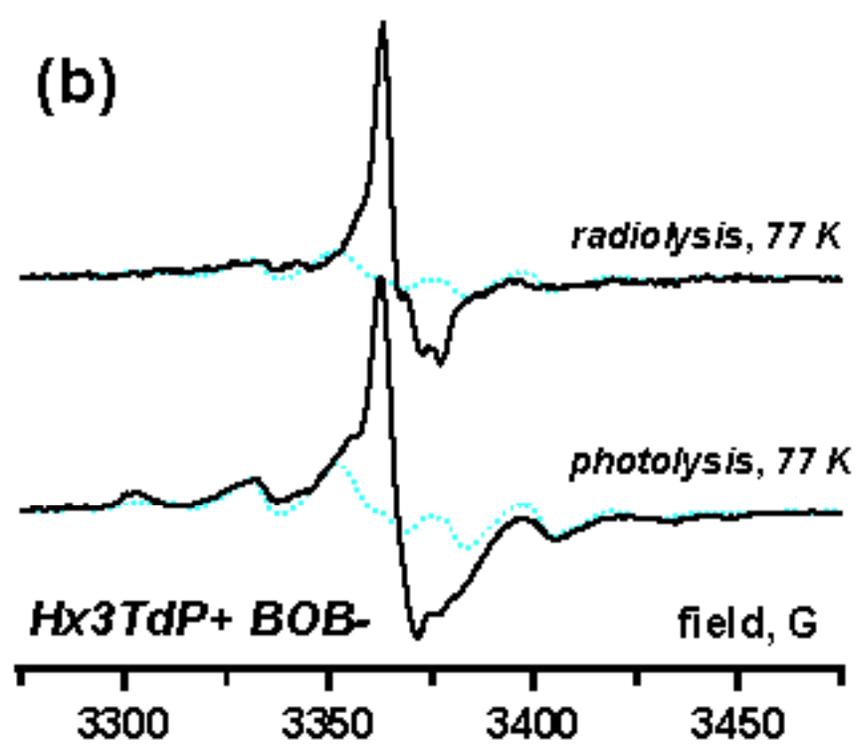

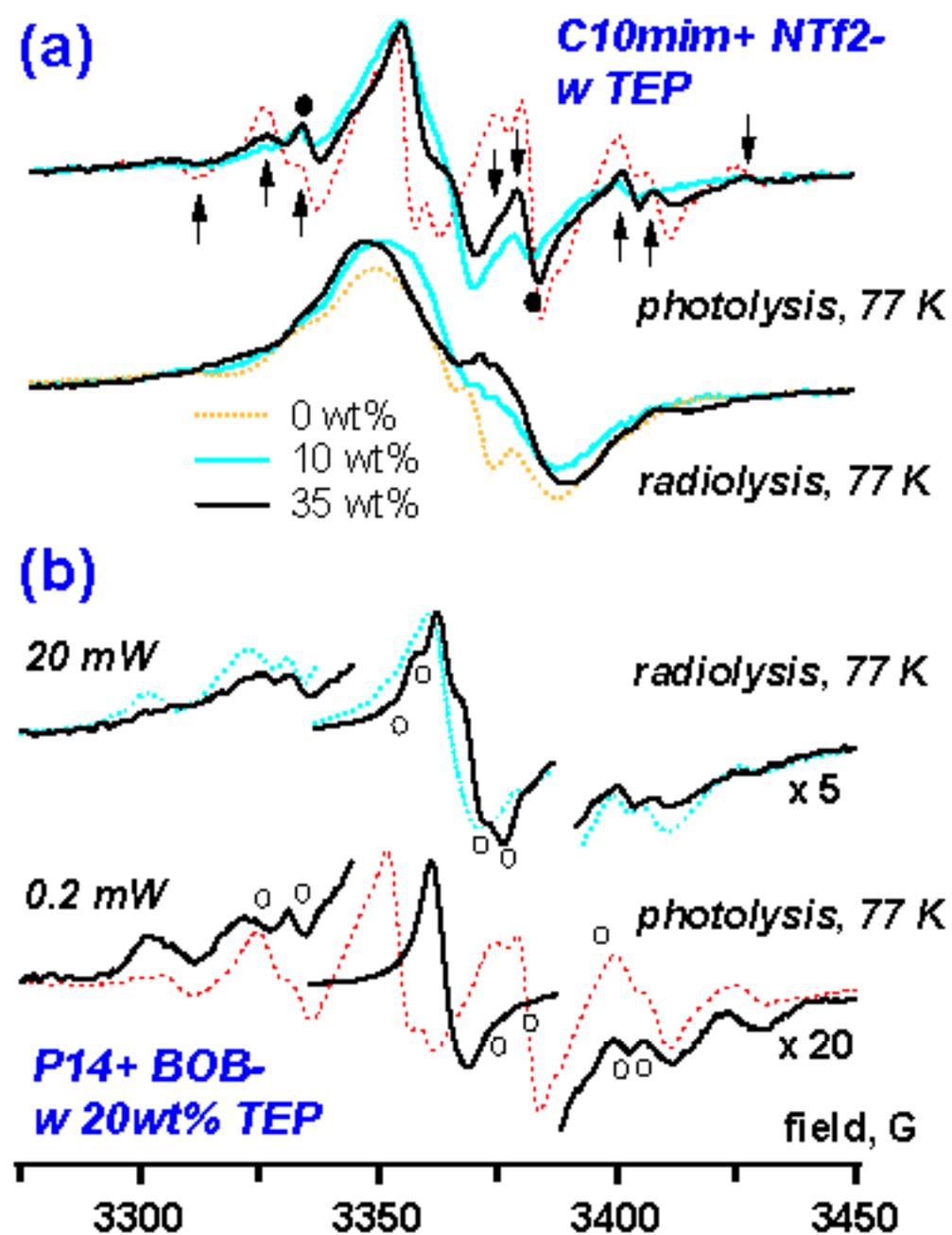

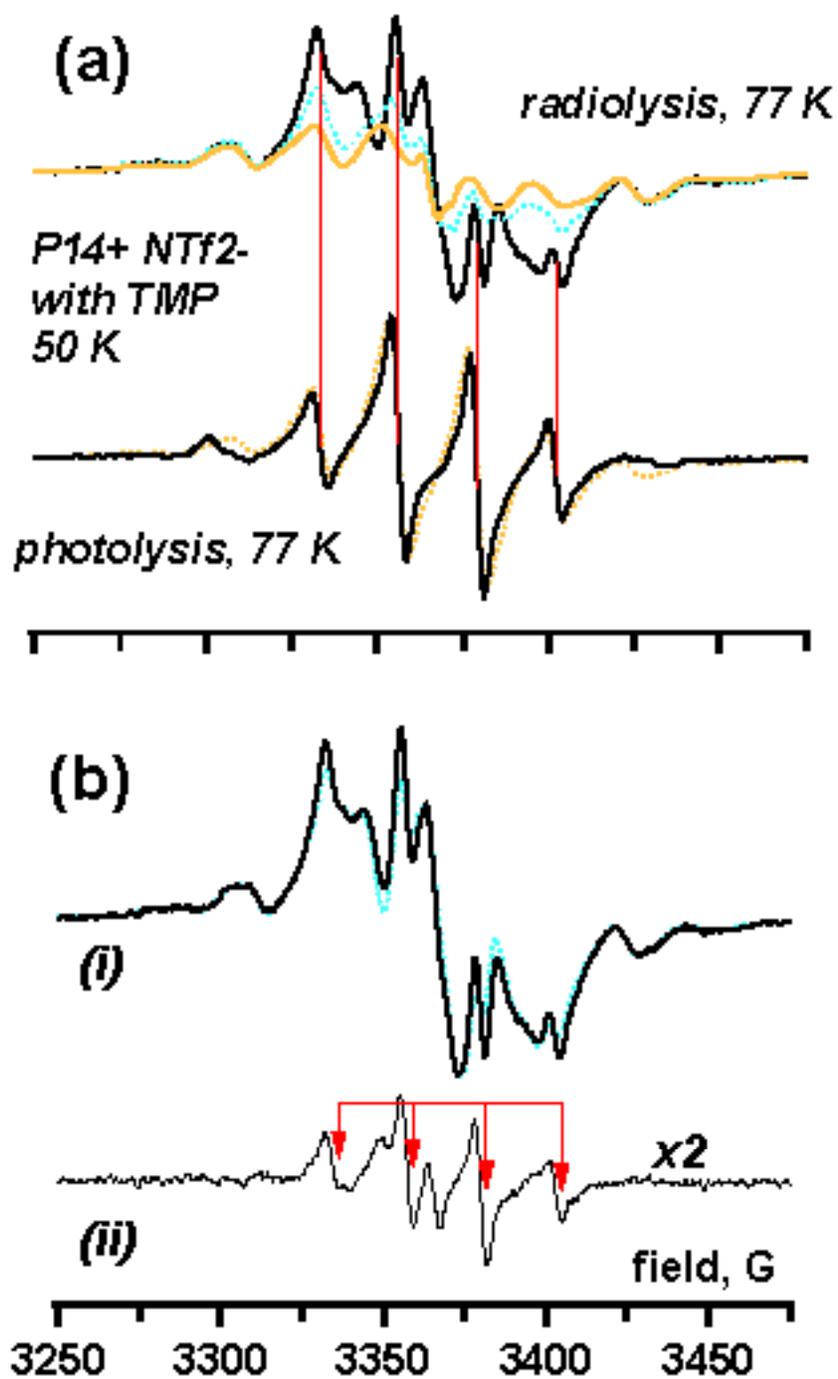

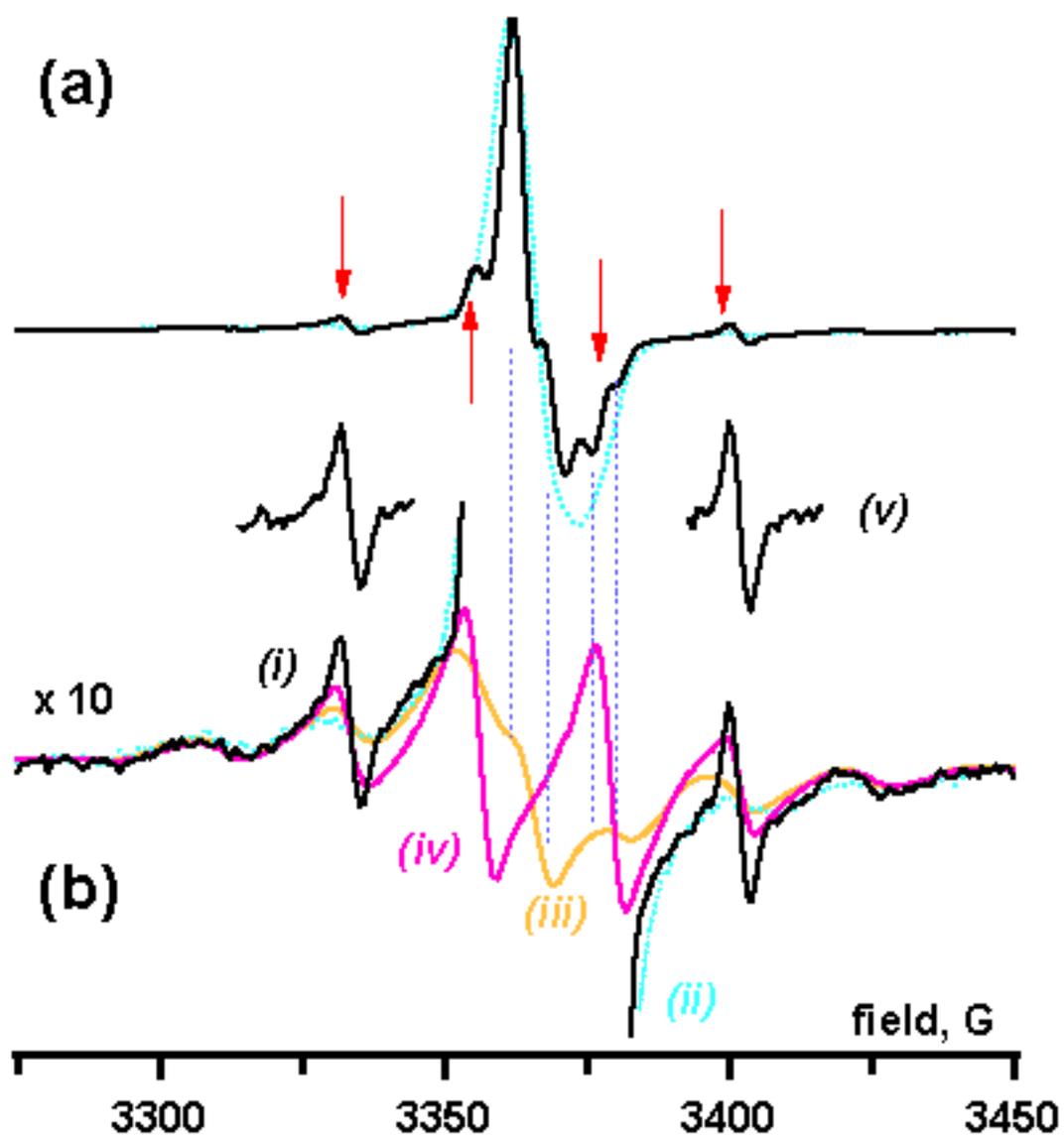